\documentclass[english]{article}
\usepackage[sc]{mathpazo}
\usepackage[LGR,T1]{fontenc}
\usepackage[a4paper]{geometry}
\usepackage{fancyhdr}
\usepackage{array}
\usepackage{float}
\usepackage{textcomp}
\usepackage{mathrsfs}
\usepackage{multirow}
\usepackage{titling}
\usepackage[affil-it]{authblk}
\usepackage{amsmath}
\usepackage{amssymb}
\usepackage{cancel}
\usepackage{graphicx}
\usepackage{setspace}
\usepackage[colorlinks=true, linkcolor=blue, urlcolor=blue,citecolor=blue]{hyperref}
\usepackage{comment}
\usepackage[normalem]{ulem}
\usepackage[english]{babel}
\usepackage[utf8]{inputenc}
\usepackage{csquotes}

\usepackage[backend=biber,style=authoryear-ibid,sorting=nyt,uniquelist=false,uniquename=false,maxbibnames=100,maxcitenames=2,doi=false,isbn=false,url=false,eprint=false]{biblatex}

\renewbibmacro{in:}{}

\DeclareFieldFormat[article]{pages}{#1}

\DeclareFieldFormat*{title}{#1}

\newbibmacro{string+doi}[1]{%
  \iffieldundef{doi}{#1}{\href{http://dx.doi.org/\thefield{doi}}{#1}}}
\DeclareFieldFormat{title}{\usebibmacro{string+doi}{\mkbibemph{#1}}}
\DeclareFieldFormat[article]{title}{\usebibmacro{string+doi}{#1}}

\AtEveryBibitem{\clearfield{issue}}
\AtEveryBibitem{\clearfield{number}}

\makeatletter

\ProvideTextCommand{\~}{LGR}[1]{\char126#1}

\usepackage{caption}
\usepackage{tocloft}

\usepackage{fancyhdr}
\date{}

\makeatother

\begin{document}

\title{\noindent Spintronics in 2D graphene-based van der Waals heterostructures}
\author{\noindent  David T. S. Perkins and Aires Ferreira}
\affil{School of Physics, Engineering and Technology and York Centre for Quantum Technologies, University of York, York, YO10 5DD, UK}

\begin{titlingpage}
    \maketitle
    \begin{abstract}
        \noindent{Spintronics has become a broad and important research field that intersects with magnetism, nano-electronics, and materials science. Its overarching aim is to provide a fundamental understanding of spin-dependent phenomena in solid-state systems that can enable a new generation of spin-based logic devices. Over the past decade, graphene and related 2D van der Waals crystals have taken center stage in expanding the scope and potential of spintronic materials. Their distinctive electronic properties and atomically thin nature have opened new opportunities to probe and manipulate internal electronic degrees of freedom. Purely electrical control over conduction-electron spins can be attained in graphene-transition metal dichalcogenide heterostructures, due to proximity effects combined with graphene's high electronic mobility. Specifically, graphene experiences a proximity-induced spin-orbit coupling that enables efficient spin-charge interconversion processes; the two most well-known and at the forefront of current research are the spin Hall and inverse spin galvanic effects, wherein an electrical current yields a spin current and non-equilibrium spin polarization, respectively. This article provides an overview of the basic principles, theory, and experimental methods underpinning the nascent field of 2D material-based spintronics.}
    \end{abstract}
\end{titlingpage}

\lhead{\noindent Spintronics in 2D graphene-based van der Waals heterostructures}

\section*{Acronym Glossary}

\noindent\textbf{2DEG}: two-dimensional electron gas

\noindent\textbf{BR}: Bychkov-Rashba

\noindent\textbf{CEE}: collinear Edelstein effect

\noindent\textbf{DOF}: degree of freedom

\noindent\textbf{FM}: ferromagnet or ferromagnetic

\noindent\textbf{ISGE}: inverse spin galvanic effect

\noindent\textbf{ISHE}: inverse spin Hall effect

\noindent\textbf{hBN}: hexagonal boron nitride

\noindent\textbf{KM}: Kane-Mele

\noindent\textbf{SdH}: Shubnikov-de Hass

\noindent\textbf{SGE}: spin galvanic effect

\noindent\textbf{SHE}: spin Hall effect

\noindent\textbf{SOC}: spin-orbit coupling

\noindent\textbf{SOT}: spin-orbit torque

\noindent\textbf{SV}: spin-valley

\noindent\textbf{TB}: tight binding

\noindent\textbf{TMD}: transition metal dichalcogenide

\noindent\textbf{vdW}: van der Waals

\noindent\textbf{WAL}: weak localization

\noindent\textbf{WL}: weak anti-localization

\newpage

\section*{Introduction\label{sec:Introduction}}

\noindent Spintronics can be considered the magnetic counterpart to electronics whereby the transfer and processing of information can be conducted through the electron's spin degree of freedom, rather than or in addition to its electronic property of charge. Specifically, spintronics concerns itself with the electrical manipulation of the electron's spin and one of its first milestones was the observation of a spin-polarized current by Tedrow and Meservey in 1973 \parencite{Tedrow1973}. Despite this and the first controlled injection of a spin-polarized current into a non-magnetic material being achieved in 1985 \parencite{Johnson1985}, it was only in 1988 that spintronics began to bloom and flourish into the field we know today. What sparked this scientific boom was the independent observation of a giant magnetoresistance attributed to the relative orientation of the ferromagnetic (FM) layers in Fe-Cr-based systems by \parencite{Baibich1988,Binasch1989}.

Since the Nobel prize winning discovery of giant magnetoresistance, initial efforts in the field focused on spin valve setups to study how the electron's spin would diffuse and relax through a normal metal when injected via a spin polarised current using an FM contact. However, a more exotic method of spin current generation had already been proposed in 1971 by Dyakonov and Perel \parencite{Dyakonov1971}, the \textit{spin Hall effect} (SHE), where the application of an electrical current would yield a perpendicular pure spin current without the need for magnetic materials. The origin of this effect can be found in the asymmetric scattering of electrons based upon their spin due to the presence of spin-orbit coupling. Initially observed in 2004 via optical methods \parencite{Kato2004}, the SHE and its inverse (ISHE) have become paradigmatic phenomena in spintronics due to their lack of reliance upon magnetic components. A major focus for the application of the SHE is in low-power magnetic memory devices, where the manifestation of a spin current results in a spin accumulation, and hence a net spin polarization, which can exert a torque on the magnetization of a nearby FM. With a large enough spin accumulation, dramatic effects like magnetization switching can be induced. Clearly, with the use of a simple electrical current, we can manipulate the spin of charge carriers in a material and induce interesting nonequilibrium phenomena at interfaces between materials, thus yielding important applications in modern information technology.

In addition to spin current, another major concept in spintronics is \textit{spin texture}: the momentum dependence of the transport electrons' spin within a solid. In many low-dimensional systems and at heterostructure interfaces, the enhancement of the relativistic spin-orbit interaction is key to generating non-trivial spin textures that give rise to a plethora of phenomena, ranging from the spin-momentum-locked surface states in topological insulators to the topologically protected real-space spin textures seen in chiral magnets, such as skyrmions and spin spirals. Of particular interest is the emergence of a spin polarization, as in the SHE, but without a resulting spin current. This effect, the sole spontaneous generation of a spin polarization via application of an electric current, is known as the \textit{inverse spin galvanic effect} (ISGE), though in some literature it may also be referred to as the Rashba-Edelstein or Edelstein effect. Based upon Bychkov-Rashba-type spin-orbit coupling (SOC) \parencite{Bychkov1984}, arising when the mirror symmetry about a given plane is broken, this effect was initially predicted in two-dimensional electron gases (2DEGs) in semiconductor heterostructures in 1989 \parencite{Aronov1989,Edelstein1990,Aronov1991} but remained unobserved in experiment until 2002, when its reciprocal effect (i.e. spin-to-charge conversion) was observed \parencite{Ganichev2002}. Combining both the SHE and ISGE, we find ourselves in a position with great control over the motion and net orientation of the electron spins in materials. However, to truly construct realistic devices using a spin-focused infrastructure, we must understand what makes an ideal material for providing such intricate control over spin. The two most important factors governing the effectiveness of spintronic devices are disorder and SOC: both mechanisms yield significant consequences on the transport range of electron spins and our control over them. While a larger SOC might ensure more efficient generation of spin currents and polarizations, it comes at the cost of faster spin dephasing and hence spin information is lost over a much shorter distance. In contrast, a weak SOC might allow for long range transport, but results in less efficient charge-to-spin conversion. Similarly, disorder can also change the efficiency of spin-charge interconversion. In a pristine system, the only processes present will be those intrinsic to the system. However, upon the inclusion of disorder, some intrinsic mechanisms are completely suppressed in favor of extrinsic mechanisms. Furthermore, spin-orbit active impurities can have similar effects to including a SOC into the system, due to their ability to change an electron's spin orientation.  Clearly, there is a balancing act to be handled in constructing ideal spintronic devices with both disorder and SOC acting as the tuning knobs.

Prior to 2004, experiments studying spin transport focused primarily on 2DEGs realized in a multitude of systems including thin metallic films \parencite{Jedema2001}, permalloys \parencite{Steenwyk1997,Dubois1999}, and semiconductor heterostructures \parencite{Yang1994,Kikkawa1999,Ohno1999,Malajovich2000}. In most cases, spin information was passed into the system by either a spin polarized charge current, or via optical excitation of a semiconductor resulting in a spin imbalance of the conduction band. Despite these initial endeavors, the electrical processing of spin-encoded information was hindered by the difficulty of combining effective spin control with large enough spin lifetimes. However, with the discovery of graphene in 2004 as the first truly 2D, i.e. atomically thin, solid state system, a new epoch dawned in spintronics. It was quickly established that bare graphene offered the largest spin diffusion lengths of any material to date, with many reports finding $l_{s} \sim 1 - 20 \, \mu$m, courtesy of the weak intrinsic spin-orbit and hyperfine interactions of $sp^{2}$ hybridized carbon.

Naturally, graphene's extremely small intrinsic SOC makes it difficult to have precise electrical control over the net spin orientation, but does allow for long-distance spin-information transfer. However, with the many advances in nanofabrication over the past two decades and the isolation of other 2D materials, such as transition metal dichalcogenides (TMDs), layer-by-layer assemblies of 2D materials have been imagined and constructed with the ability to alter the net behavior of the composite system based purely on the individual layers used. Thus the concept of bespoke devices combining the desirable properties of different materials has become possible in the form of van der Waals (vdW) heterostructures.

What single 2D crystals lack, vdW heterostructures may offer a way to access by enhancing certain properties by matching an appropriate set of materials. Through simple proximity effects, the properties absent or deemed too weak in the original isolated crystal can be enhanced. Semiconducting TMDs, such as MoS${}_{2}$ and WTe${}_{2}$, are a classic example of this; due to their composition involving transition metal atoms, they naturally possess a large SOC. By stacking this with a graphene monolayer, the electrons of the graphene sheet will experience a \textit{proximity-induced SOC effect}, due to their ability to now hop between the two layers. Furthermore, graphene's transport nature is not jeopardised by the proximity of the TMD since the low-energy states of graphene lie well within the TMD band gap. Consequently, transport is still dominated by the more conductive graphene layer, though the electronic states are now endowed with a substantial SOC. The typical size of these induced SOCs is up to order 10 meV \parencite{Wang2019}, which is three orders of magnitude larger than the intrinsic SOC present in regular graphene (Kane-Mele-type of order 10 $\mu$eV \parencite{Sichau_19}). The specific values of the SOCs depends on the choice of TMD partner and can be further tuned by means of external pressure \parencite{Fulop2021}. This dramatic change from an isolated crystal's properties with the introduction of material partners is what makes vdW heterostructures the modern candidates par excellence for many condensed matter experiments.

\section*{Graphene and van der Waals Heterostructures}

\begin{figure}[t]
    \begin{centering}
    \includegraphics[clip,width=1\textwidth]{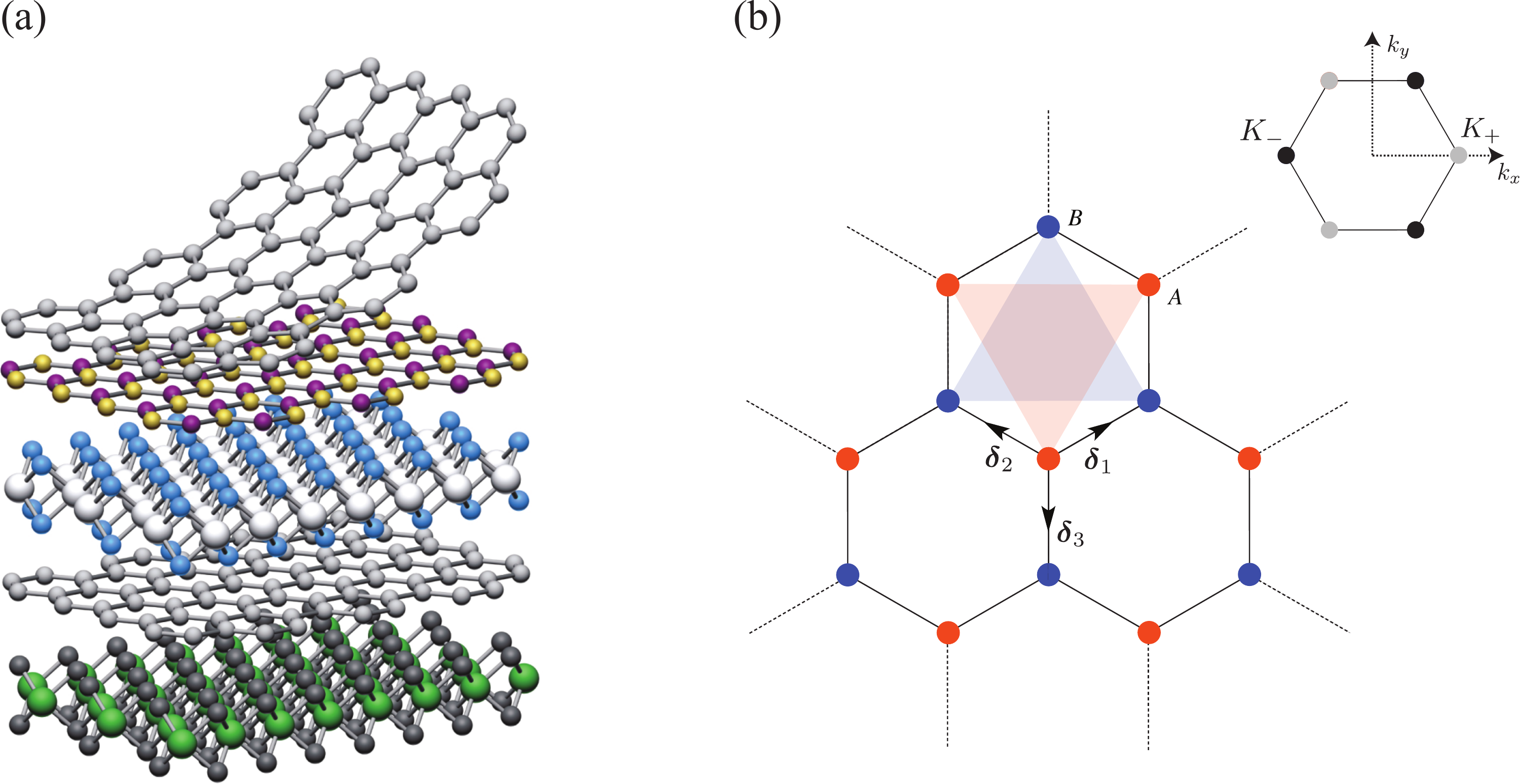}
    \par\end{centering}
    \caption{\label{fig:01}(a) Vertical heterostructure built from 2D crystals. From Ref.~\parencite{Geim2013} with permission from the authors. (b) The honeycomb lattice composed of two triangular Bravais lattices (top right corner shows the first Brillouin zone with two inequivalent Dirac points, $K_{\pm}$). The nearest-neighbor vectors read as $\boldsymbol{\delta}_{i}=a\,(\sin\alpha_{i},-\cos\alpha_{i})^{\textrm{T}}$, where $a$ is the bond length and $\alpha_{i}=\frac{2\pi}{3}(i-3)$ and $i$=1,2,3. }
\end{figure}

Formed by stacking atomically thin layers of hexagonally packed atoms with weak van der Waals forces binding neighboring layers together, Fig. \ref{fig:01}a \parencite{Geim2013}, vdW heterostructures constitute a specific class of 2D materials. Given the breadth of materials with 2D behavior -- encompassing semiconductors, insulators, and semimetals -- which can be exfoliated down to a monolayer, it is no surprise that the variety of vdW heterostructures is equally diverse. Furthermore, the electronic properties of such 2D compounds are sensitive to the number of layers, stacking sequence, and atomic coordination, while also being tunable ``on-demand'' through the controlled application of strain and electric fields \parencite{Castro_Neto2009,Wang2012,Yun2012}. 

Monolayer graphene presents itself as a zero-gap semiconductor with a linear dispersion relation for both electrons and holes, whose unit cell consists of two distinct lattice sites which are individually referred to as the A and B sublattices (Fig. \ref{fig:01}b). The band structure of graphene is characteristic of \textit{massless chiral Dirac fermions}, which derives from the $sp^{2}$ network of carbon atoms forming a honeycomb lattice with preserved inversion symmetry (Fig. \ref{fig:01}b), whose presence give rise to graphene's notable electronic properties \parencite{Castro_Neto2009}. Other 2D materials relevant to vdW heterostructures include: (i) hexagonal boron nitride ``hBN'', an insulator, (ii) bilayer graphene, a zero-gap semiconductor with parabolic band dispersion; and (iii) group-VI dichalcogenides, direct band-gap semiconductors with strong SOC.  

What makes a material useful in vdW heterostructures varies from material to material. The band gap opening in polyatomic compounds (e.g. hBN) is a direct consequence of broken inversion symmetry. The electronic band gap in monolayer hBN is about 7 eV, making it an insulating analog of graphene. The small lattice mismatch between hBN and graphene (about $1.8 \%$) allows for easy integration in graphene-based devices using dry transfer techniques. hBN encapsulation yields major improvements in the electronic mobility of graphene-based devices, and is currently the gold standard for the fabrication of high-performance vdW heterostructures. In lateral spin transport experiments, the hBN encpasulation of graphene has enabled a 10-fold increase in room-temperature spin lifetimes compared to devices using the now obsolete silicon oxide substrates \parencite{Drogeler2014}. The hBN induces an orbital gap in graphene (Fig. \ref{fig:02} (a)), which results from virtual interlayer hopping processes between carbon sites in graphene and the distinct chemical species occupying the A and B sublattices of the partnered hBN layer.

Bilayer graphene can exist in Bernal-stacked ``AB'' form, with atoms on opposite layers stacked on top of each other in a staggered configuration, or, less frequently, in the ``AA'' form, with sublattices in adjacent layers perfectly aligned \parencite{Liu2009}. Similar to monolayer graphene, its conduction and valence bands touch at the Brillouin zone corners (Fig.~\ref{fig:02} (a)), however low-energy excitations around the Fermi points are associated with quadratic energy dispersion \parencite{McCann2006a,Guinea2006}. The finite density of states near the Fermi level exacerbates the effect of interactions leading to rich broken symmetry states even in the absence of external fields \parencite{Zhang2010,Vafek2010,Nandkishore2010,Weitz2010}. Individual layers in AB bilayer graphene can be addressed separately allowing important device functionalities, including band gap opening through gating \parencite{McCann2006b,Castro2007}. Of particular interest for spintronics, is the ability to fine-tune both the SOC (proximity with a TMD) and exchange interaction (proximity with an FM) experienced by the electrons. Only the layer that is the immediate neighbor to a partner material will experience significant proximity effects. Therefore, by applying an electric field perpendicular to the graphene bilayer, the electron density of each layer can be adjusted, which in turn changes the proximity-induced SOC and exchange interaction experienced by the electrons by shifting them towards or away from the partner material \parencite{Zollner2020}.

Group-VI  dichalcogenides {[}MX$_{2}$ (M$=$Mo, W; X$=$S, Se,Te){]} can be either trigonal-prismatically or octahedrally coordinated (so-called ``2H'' and ``1T'' phases, respectively). These polytypic structures have vastly different electronic properties: while TMDs of the 2H-MX$_{2}$ type are large-gap semiconductors, 1T-MX$_{2}$ are predominantly metals \parencite{Mattheiss1973,Eda2012,Voiry2015}.  2H-TMD monolayers have direct band gaps in the near-infrared to the visible region, which make them well suited for a broad range of applications in optoelectronics and photonics \parencite{Mak2016}. Owing to ultimate (2D) quantum confinement, electrons and holes are tightly held together which is reponsible for enhanced light--matter interactions. Excitons have typical binding energies of 0.5 eV, which strongly impact the spin and optoelectronic properties of semiconducting TMDs \parencite{Wang2018}.  Moreover, the spin-valley locking of energy states in the vicinity of $K_{\pm}$ points, stemming from lack of inversion symmetry, lead to spin-valley-dependent optical transition rules \parencite{Wang2012}, which enable the addressing of individual valleys with circularly polarized light. The spin\textminus valley coupling in TMDs has been explored to convert optically driven valley currents into charge currents via the inverse valley Hall effect \parencite{Mak2014}. In free-standing conditions, the 1T phase undergoes a spontaneous lattice distortion to a semiconducting phase dubbed 1T$^{\prime}$, which supports robust nontrivial topological behavior (quantum spin Hall effect) \parencite{Qian2014,Tang2017,Wu2018,Shi_19}. 

The weak van der Waals forces between planes of 2D crystals offer a pratical route for band structure design. A remarkable example is ``twisted'' bilayer graphene: where the two graphene monolayers forming bilayer graphene are offset from one another by a simple rotation about the out-of-plane axis \parencite{Lopes_dos_Santos2007}, in addition to their stacking arrangement. For ``magic angle'' twisted bilayer graphene, the superlattice created by the two graphene sheets leads to strong renormalization of the band structure, which can exhibit flat bands at the Fermi level, leading to possible strongly correlated insulating states in the ultimate 2D (atomically thin) limit \parencite{Cao2018}.

With the above discussion of different 2D systems and their various combinations, it is clear that van der Waals heterostructures offer a route towards 2D designer materials. The exact combination of the individual layers is dictated by the purpose of the device being constructed. This is particularly relevant when trying to study and apply the SHE and ISGE from a technological perspective. The specifics behind these phenomena are discussed towards the end of this article, while the effect of proximity-induced SOC shall be covered shortly.

\subsection*{Honeycomb monolayers: \emph{A Tight-binding Description}}

The distinctive electronic properties of 2D layered materials and the special role played by the sublattice degree of freedom (DOF) can be best appreciated within a tight-binding (TB) model of electrons hopping on a honeycomb lattice (Fig.~\ref{fig:01} (b)). The minimal Hamiltonian (without SOC) reads as
\begin{figure}
    \begin{centering}
    \includegraphics[clip,width=1\textwidth]{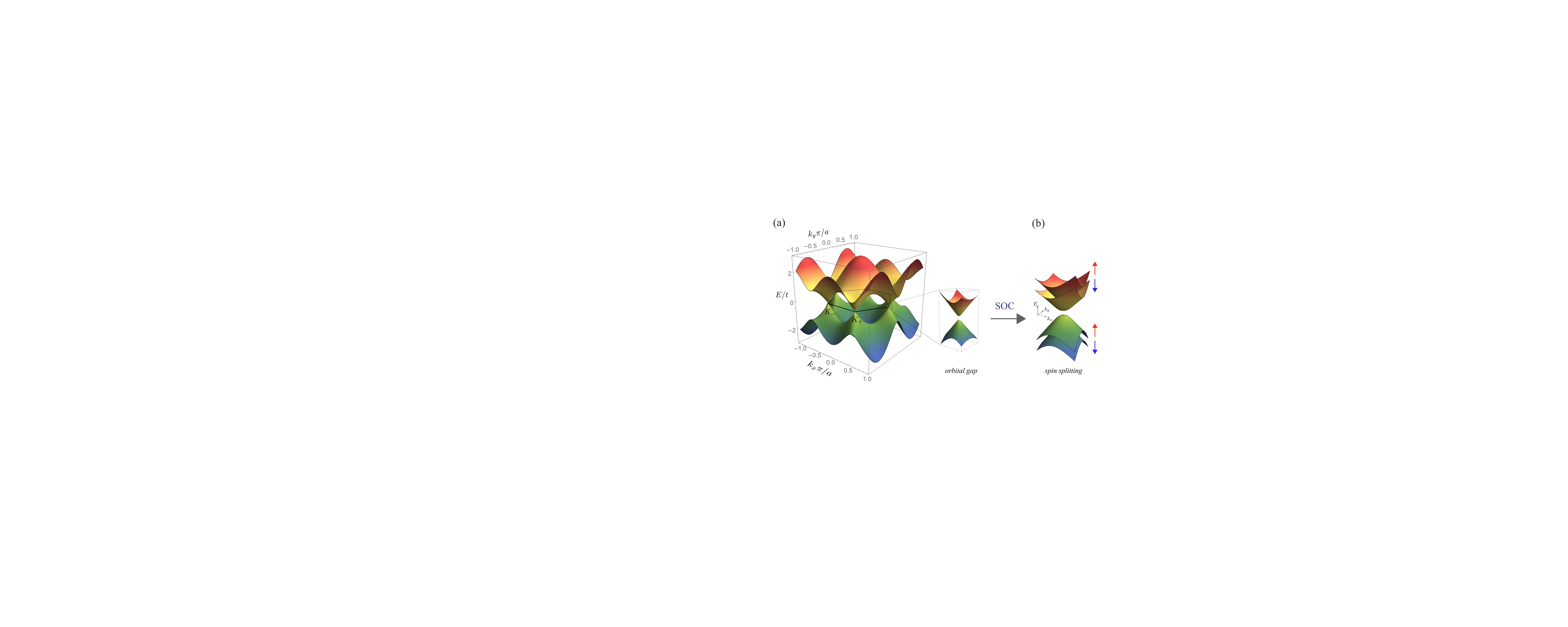}
    \par\end{centering}
    \centering{}\caption{\label{fig:02}(a) Low-energy dispersion of a honeycomb monolayer with a small orbital gap ($m\protect\neq0$). (b) Lifting of spin degeneracy due to symmetry breaking SOC (arrows indicate different spin states).}
\end{figure}
 
\begin{equation}
    H_{\textrm{2D}} = -t\sum_{\langle i,j\rangle} \left( a_{i}^{\dagger}b_{j} + \textrm{H.c.} \right) + m \sum_{i} \left( a_{i}^{\dagger}a_{i} -b_{i}^{\dagger}b_{i} \right)\, 
    \label{eq:tight_binding_nn}
\end{equation}
where the fermionic operator $a_{i} \, (b_{i})$ annihilates a quasiparticle on site $i$ belonging to sublattice $A$ ($B)$, $t$ is the nearest neighbor hopping energy ($t\approx2.8$~eV in graphene \parencite{Castro_Neto2009}), and $\langle i,j \rangle$ denotes the sum over nearest neighbors. The second term describes a staggered on-site energy with amplitude $m=(\varepsilon_{A}-\varepsilon_{B})/2$ and is relevant for noncentrosymmetric 2D crystals (e.g. $m\approx 3.5$ eV in hBN \parencite{Roman_2021} and $m\approx 0.7-0.9$ eV in semiconducting TMDs \parencite{Xiao_2012}), as well as graphene-based heterostructures displaying moir\'{e} superlattice effects \parencite{Dean2010,Jung2015,Wallbank_15}.

This tight-binding Hamiltonian provides a starting point for understanding the low-energy properties of prototypical 2D materials. The energy bands are obtained by means of the Fourier transform $a_{i} \, (b_{i})=N^{-1}\sum_{\boldsymbol{k}}e^{-i\boldsymbol{k}\cdot\boldsymbol{x}_{i}}a_{\boldsymbol{k}} \, (b_{\boldsymbol{k}})$, where $\boldsymbol{k}=(k_{x},k_{y})^{\textrm{T}}$ is the 2D wavevector and $N$ is the number of sites in each sublattice ($N=N_{A}=N_{B}$). After straightforward algebra, one finds
\begin{equation}
    H_{\textrm{2D}} = \sum_{\boldsymbol{k}} \begin{pmatrix}
        a_{\boldsymbol{k}}^{\dagger} & b_{\boldsymbol{k}}^{\dagger}
    \end{pmatrix}
    \begin{pmatrix}
        m & \phi^{*}(\boldsymbol{k})\\
        \phi(\boldsymbol{k}) & -m
    \end{pmatrix}
    \begin{pmatrix}
        a_{\boldsymbol{k}}\\
        b_{\boldsymbol{k}}
    \end{pmatrix},
    \label{eq:Ham_TB_momentum_space}
\end{equation}
with the geometric form factor 
\begin{equation}
    \phi(\boldsymbol{k}) = -t \sum_{a=1,2,3} e^{i\boldsymbol{k} \cdot \boldsymbol{\delta}_{a}}\,,
    \label{eq:geometric_form_factor}
\end{equation}
where the bond vectors, $\boldsymbol{\delta}_{a}$, are defined in the caption of Fig.~\ref{fig:01}. The energy dispersion is readily obtained as
\begin{equation}
    E(\boldsymbol{k}) = \pm \sqrt{t^{2} |\phi(\boldsymbol{k})|^{2} + m^{2}}\,,
    \label{eq:energy_spectrum}
\end{equation}
where the sign $\pm$ selects positive (negative) energy branch of the spectrum. A band structure representative of graphene with a sublattice staggered potential is shown in Fig. \ref{fig:02}(a). Of particular note is the appearance of local extrema in the spectrum at the corners of the Brillouin zone, $K_{\pm}$, where $E=\pm m$, known as Dirac points. In this example, the small orbital gap ($m\ll t$) could be due to to use of a lattice-matching substrate (e.g. hBN). At half-filling, the negative energy states are completely filled, and hence the low-energy physics is controlled by excitations about the Dirac points (the region around a Dirac point is known as a valley). For semiconducting TMDs and hBN, the staggered on-site energy is comparable to (or larger than) $t$, resulting in sizable orbital gaps at the Dirac points. The generalization of this Hamiltonian to describe the strong intrinsic SOCs inherent in semiconducting TMDs, as well as symmetry breaking SOCs in vdW heterostructures, is presented in the next section. while the extension of the effective TB model to  multilayers (e.g. bilayer graphene) is straightforward, the complete details are beyond the scope of this article, though the interested readers can find details in Refs. \parencite{Peres2010,McCann2013}.

\subsection*{Honeycomb monolayers: \emph{SOC interactions}}

The electronic structure of 2D materials containing heavy elements is strongly modified by spin-orbit effects generated by the periodic crystal potential. In the nonrelativistic approximation to the Dirac equation, the intrinsic spin-orbit interaction reads as
\begin{equation}
    H_{\text{SO}} = -\frac{\hbar}{4m_{e}^{2}c^{2}} \, \boldsymbol{s} \cdot (\boldsymbol{p} \times \boldsymbol{\nabla}V)\,,
    \label{eq:SO_nonrelativistic}
\end{equation}
where $V(\boldsymbol{x})$ is the periodic crystal potential, $\boldsymbol{p}$ is the momentum operator, $m_e$ is the electron mass, and $\mathbf{s}$ is the vector of Pauli matrices describing spin-$1/2$ particles (i.e. acting on the spin DOF). The (spin-dependent) hoppings generated by Eq.~(\ref{eq:SO_nonrelativistic}) can be obtained by exploring time-reversal symmetry ($\mathcal{T}$) and the crystal symmetries. The most general $H_{\text{SO}}$ for honeycomb lattices may therefore be written as  
\begin{equation}
    H_{\textrm{SO}}^{\mathcal{T}} = \sum_{\langle i,j\rangle} \left( \hat{a}_{i}^{\dagger} {T}_{ij}^{ab} \hat{b}_{j} + \hat{b}_{i}^{\dagger} {T}_{ij}^{ba} \hat{a}_{j} \right) \, + \sum_{\langle\langle i,j\rangle\rangle} \left( \hat{a}_{i}^{\dagger} {T}_{ij}^{aa} \hat{a}_{j} + \hat{b}_{i}^{\dagger} {T}_{ij}^{bb} \hat{b}_{j} \right)\,,
    \label{eq:TB_SO_general}
\end{equation}
where $\hat{a}_{i}^{\dagger}$/$\hat{a}_{i}^{\null}$ ($\hat{b}_{j}^{\dagger}$/$\hat{b}_{j}^{\null}$ ) are creation/annihilation operators for the $A$($B$) sublattice with a 2-component spinor structure acting on spin space, ${T}_{ij}^{(\varsigma)}$ (with $\varsigma=aa,ab,bb$) are spin-dependent hopping coefficients with a $2\times2$ complex matrix structure satisfying ${T}_{ij}^{\varsigma}=[{T}_{ji}^{\varsigma}]^{\dagger}$, and $\langle\langle i,j \rangle\rangle$ denotes the sum over next-nearest neighbors. The neglect of hoppings beyond next-nearest neighbors in Eq.~(\ref{eq:TB_SO_general}) is justified  given the exponential decay of matrix elements with the distance. Note that on-site spin-dependent terms, such as $\hat{a}_{i}^{\dagger}s_{z}\hat{a}_{i}$, change sign under $\mathcal{T}$ and thus are not allowed.
\begin{figure}
    \begin{centering}
    \includegraphics[width=0.8\textwidth]{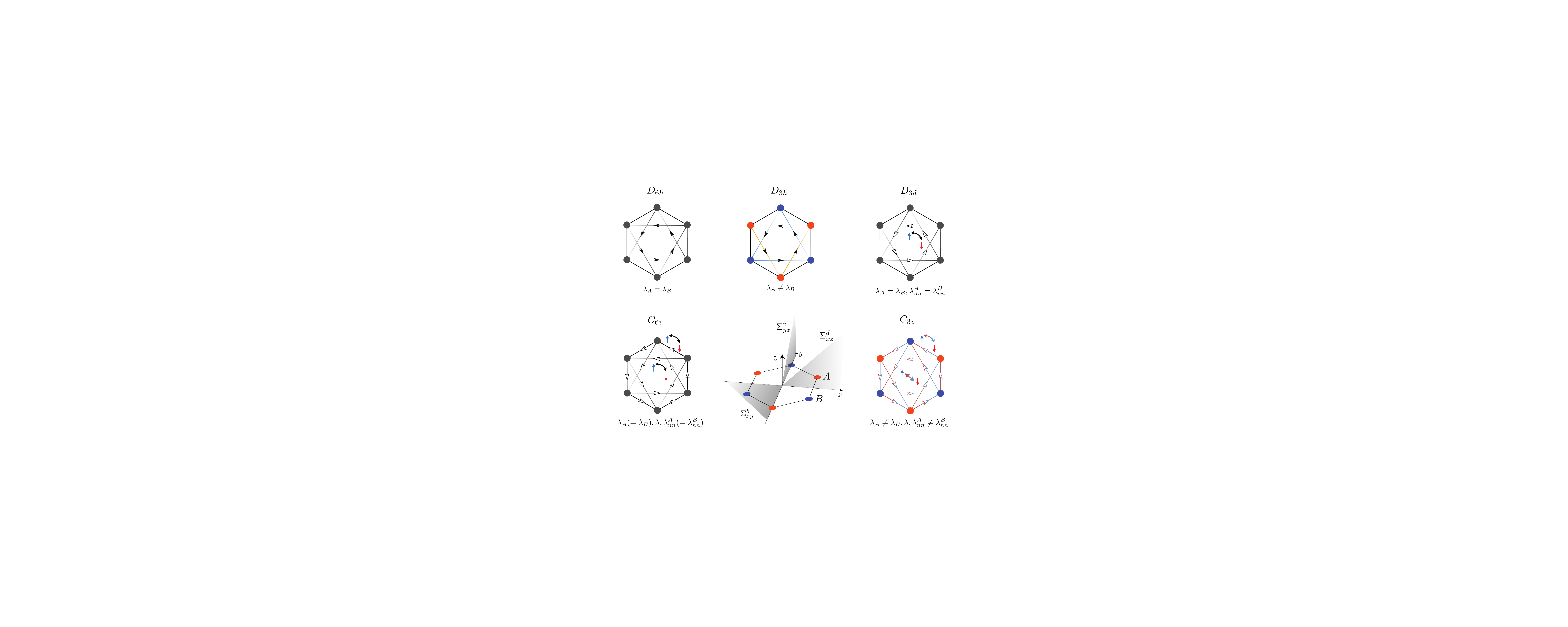}
    \par\end{centering}
    \caption{\label{fig:03}Spin--orbit hopping terms compatible with subgroups of $D_{6h}$. Horizontal ($h$), vertical ($v$) and dihedral ($d$) reflections are shown (bottom, middle). Black (white) arrows indicate spin-conserving (spin-flip) hoppings. $D_{6h}$-, $D_{3h}$- and $D_{3d}$- invariant systems (top panel) admit next-nearest hoppings. $C_{6v}$- and $C_{3v}$- invariant systems (bottom panel, right and left) allow for nearest neighbor spin-flip hopping.}
\end{figure}

The symmetries of the system are contained within the ${T}_{ij}^{(\varsigma)}$. Expansion of the spin hopping matrices into elements of the SU(2) spin-algebra yields
\begin{equation}
    {T}_{ij} = \omega_{ij}^{x} \, s_{x} + \omega_{ij}^{y} \, s_{y} + \omega_{ij}^{z} \, s_{z} \,,
    \label{eq:spin_hopping_matrices}
\end{equation}
where sublattice superscripts have been omitted. The anti-unitary operator enacting $\mathcal{T}$ is given by $\Theta=i s_{y}K$, where $K$ denotes complex conjugation. The requirement, ${T}_{ij}=\Theta\,{T}_{ij}\,\Theta^{-1}$, leads to the following constraint: $\omega_{ij}^{\alpha} = -(\omega_{ij}^{\alpha})^{*}$ (with $\alpha=x,y,z$). As such, the coefficients of spin-dependent hopping mediated by $H_{\text{SO}}$ are purely imaginary. The full set of spin--orbit interactions up to next-nearest neighbors is shown in Fig.~\ref{fig:03}, where the allowed hoppings are dictated by the point group symmetry. For example, $D_{6h}$-invariant Hamiltonians (e.g. flat pristine graphene) only permit spin-conserving next-nearest neighbor hoppings. This is the result of three symmetries. First, mirror inversion about the plane, $\Sigma_{xy}^{h}$, which reverses the sign of $x$- and $y$- components in Eq.~(\ref{eq:spin_hopping_matrices}), forbids all spin flip processes, $\omega_{ij}^{(x,y)}=0$. Second, the mirror symmetry $\Sigma_{xz}^{d}$, which sets $y \rightarrow -y$ and consequently reverses the sign of $s_{z}$ in Eq. (\ref{eq:spin_hopping_matrices}), ensures that $\omega_{ij}^{z} = 0$ for all in-plane vertical hoppings, which are naturally nearest neighbor processes. Finally, combining the $\Sigma_{xz}^{d}$ mirror symmetry with 6-fold rotational symmetry about the $z$-axis, we see that the vertical hoppings must map onto the other nearest neighbor hoppings and hence they too must vanish.

More generally, when space inversion or horizontal reflection symmetries are broken, other terms are allowed. Of particular relevance is the $C_{3v}$ point group, given that it is a common subgroup of $D_{3h}$ (e.g. hBN and TMD monolayers), $D_{3d}$ (e.g. rippled graphene and Si- and Ge-based graphene analogs), and $C_{6v}$ (e.g. graphene on a non-lattice-matched substrate) and hence defines the most general class of Hamiltonians compatible with the honeycomb lattice symmetries \parencite{Saito_2016,Kochan2017}. Altogether, $C_{3v}$-invariant models allow for 3 types of SOC  
\begin{equation}
\begin{split}
    H_{\textrm{SO}}^{C_{3v}} & = \frac{i}{3\sqrt{3}} \sum_{\langle\langle i,j \rangle\rangle} \nu_{ij} \left( \lambda_{A} \, \hat{a}_{i}^{\dagger}s_{z}\hat{a}_{j} + \lambda_{B} \, \hat{b}_{i}^{\dagger}s_{z}\hat{b}_{j} \right) + \frac{2i}{3} \sum_{\langle i,j\rangle} \left[ \lambda \, \hat{a}_{i}^{\dagger} \left( \boldsymbol{s} \times \boldsymbol{d}_{ij} \right)_{z} \hat{b}_{j}^{\null} + (b \leftrightarrow a) \right] \\
    & \;+ \frac{2i}{3} \sum_{\langle\langle i,j \rangle\rangle} \left[ \lambda_{nn}^{A} \, \hat{a}_{i}^{\dagger} \left( \boldsymbol{s} \times \boldsymbol{d}_{ij} \right)_{z} \hat{a}_{j}^{\null} + \lambda_{nn}^{B} \, \hat{b}_{i}^{\dagger} \left( \boldsymbol{s} \times  \boldsymbol{d}_{ij} \right)_{z} \hat{b}_{j}^{\null} \, \right],
    \label{eq:C3v_invariant}
\end{split}
\end{equation}
namely, spin-conserving sublattice-resolved SOC, $\lambda_{A(B)}$, Bychkov-Rashba (BR) interaction, $\lambda$, and spin-flipping sublattice-resolved SOC, $\lambda_{nn}^{A(B)}$. Here, $\boldsymbol{d}_{ij}$ is the unit vector along the line segment connecting site $i$ and $j$, while $\nu_{ij} = \pm 1$ distinguishes between clockwise and anti-clockwise electron hopping respectively.

The spin-conserving SOC (first term in Eq.~(\ref{eq:C3v_invariant})) is a fingerprint of
pseudospin-spin coupling in 2D layered materials. Two cases are of noteworthy interest: (i) centrosymmmetric crystals, where only a spin conserving SOC is permitted with $\lambda_{A} = \lambda_{B}$, as is the case in pristine graphene ($D_{6h}$) discussed above, and graphene-like materials with $D_{3d}$ point group, and (ii) systems with broken inversion symmetry leading to sublattice-resolved SOC ($\lambda_{A} \neq \lambda_{B}$), as seen, for example, in 2H-TMD monolayers ($D_{3h}$) and graphene-TMD heterostructures ($C_{3v}$). The BR interaction generated by the nearest-neighbor spin-flip hoppings (second term in Eq.~(\ref{eq:C3v_invariant}))  signals the lack of mirror inversion symmetry, $\Sigma_{xy}^{h}$, associated with reduction of the point group from $D_{6h}$ to $C_{6v}$, which may occur through interfacial effects in heterostructures or via application of an electric field perpendicular to the 2D plane. Finally, next-nearest-neighbor spin-flip processes, parameterized by the couplings $\lambda_{nn}^{A(B)}$ (third term in Eq.~(\ref{eq:C3v_invariant})) are also allowed in systems lacking out-of-plane reflection symmetry, $\Sigma_{xy}^{h}$, such as graphene-based vdW heterostructures or graphene placed on a generic substrate. Note that when the in-plane inversion symmetry is also broken, these processes become sublattice-resolved ($\lambda_{nn}^A\neq \lambda_{nn}^B$). This occurs, for example, in graphene-TMD heterostructures.

\subsection*{Honeycomb Monolayers with Proximity-Induced SOC: Continuum Theory}

The low-energy electronic structure of 2D materials can be conveniently modeled using a (long-wavelength) continuum description. Expanding the geometric form factor [Eq.~(\ref{eq:geometric_form_factor})] to first order around each Dirac point, $\boldsymbol{K}_{\pm}=4\pi/(3 \sqrt 3 a)(\pm1,0)^{\text{T}}$, that is $\phi_{\pm}(\boldsymbol{k})\simeq(3ta/2)(\pm k_{x}+ik_{y})$, yields the effective single-particle Hamiltonian $\mathcal{H}_{0\boldsymbol{k}} = \hbar v \left( \tau_{z} \otimes \sigma_{x}k_{x} + \tau_{0} \otimes \sigma_{y}k_{y} \right) \otimes s_{0} + m \tau_{0} \otimes \sigma_{z} \otimes s_{0}$ when written in the conventional basis, $|\psi\rangle = \left(K_{+}, K_{-}\right)^{\text{T}} \otimes \left(A, B\right)^{\text{T}} \otimes (\uparrow,\downarrow)^{\text{T}}$. Here $\boldsymbol{k}$ is the 2D wavevector measured with respect to a Dirac point, $v=3at/2\hbar$ is the Fermi velocity of massless Dirac fermions, $s_{0}$ is the identity matrix acting on the spin DOF, and $\sigma_{i}$ and $\tau_{i}$ are the Pauli matrices supplemented with identity acting on sublattice and valley DOFs, respectively. This low-energy Hamiltonian admits a even simpler and more elegant form when written using the \textit{valley basis}, $|\psi\rangle = \left( K_{+}A, K_{+}B, -K_{-}B, K_{-}A \right)^{\text{T}} \otimes (\uparrow,\downarrow)^{\text{T}}$, i.e. $\mathcal{H}_{0\boldsymbol{k}} = H_{0\boldsymbol{k}} \otimes s_{0}$, with
\begin{equation}
   H_{0\boldsymbol{k}} = \hbar v \, \tau_{0} \boldsymbol{\sigma} \cdot \boldsymbol{k} + m \tau_{z} \sigma_{z}.
    \label{eq:realspace_eff_Ham}
\end{equation}
The Dirac-Weyl equation [Eq.~(\ref{eq:realspace_eff_Ham})] governs the low-energy properties of honeycomb monolayers with low SOC and has been extensively used in investigations of opto-electronic and transport phenomena in single-layer graphene \parencite{Peres2010,McCann2013}. For ease of notation, the writing of the tensor product between matrices acting on different DOFs is omitted from here onward.

The continuum version of the SOC Hamiltonian ($\mathcal{H}_{\textrm{SO}}$) can be derived by expanding Eq. \ref{eq:C3v_invariant} around the $K$ points. Alternatively, akin to the tight-binding approach of the previous section, temporal and unitary (spatial) symmetries can be exploited so as to constrain the form of $\mathcal{H}_{\textrm{SO}}$. In addition to reversing momentum ($\boldsymbol{k} \rightarrow -\boldsymbol{k}$) and spin ($\boldsymbol{s} \rightarrow -\boldsymbol{s}$), $\mathcal{T}$ also reverses pseudospin ($\boldsymbol{\sigma} \rightarrow -\boldsymbol{\sigma}$) and swaps valleys (as momentum reversal sends $\boldsymbol{K}_{+} \leftrightarrow \boldsymbol{K}_{-}$). Exploiting the $\mathcal{T}$-symmetry transformations, one finds that there are $4\times3=12$ possible terms
\begin{equation}
    \mathcal{H}_{\textrm{SO}}^{\mathcal{T}} = \sum_{i=x,y,z} \: \left( \Delta^{(i)} \tau_{0} \, \sigma_{z} + \lambda_{x}^{(i)} \tau_{0} \, \sigma_{x} + \lambda_{y}^{(i)} \tau_{0} \, \sigma_{y} + \omega^{(i)} \tau_{z} \, \sigma_{0} \right) \, s_{i} \,.
    \label{eq:HSO_generic}
\end{equation}
Direct inspection of Eq. (\ref{eq:HSO_generic}) shows that the majority of SOC terms lead to anisotropic energy dispersion, and hence are incompatible with the $C_{3v}$ point group symmetry. To preserve the continuous rotational symmetry of the low-energy theory about each Dirac point, the Hamiltonian within each valley should commute with the generator of rotations within the 2D plane for that valley (the total angular momentum operator). In the valley basis, the total angular momentum is independent of the valley index and takes the form 
\begin{equation}
    J_{z} = -i \, \hbar \partial_{\phi} \tau_{0} \, \sigma_{0} \, s_{0} + \frac{\hbar}{2} \left( \tau_{0} \, \sigma_{z} \, s_{0} + \tau_{0} \, \sigma_{0} \, s_{z} \right),\label{eq:total_ang_momentum}
\end{equation}
where $\phi$ is the azimuthal angle. The first term of $J_{z}$ is simply the orbital angular momentum, while the $s_{z}$ piece is the usual spin contribution. The $\sigma_{z}$ term arises from the spin-like sublattice DOF and is therefore characteristic of vdW materials. The requirement of $J_{z}$'s commutation with the Hamiltonian yields  $\Delta^{(x,y)}=\omega^{(x,y)}=\lambda_{x(y)}^{z}=0$, $\lambda_{x}^{(y)}=-\lambda_{y}^{(x)}$ and $\lambda_{x}^{(x)}=\lambda_{y}^{(y)}$. One of such terms, a Dresselhaus-type SOC, $\lambda_{x}^{(x)} \tau_{0}(\sigma_{x} \, s_{x}+\sigma_{y} \, s_{y})$, breaks the mirror reflection symmetry about the $yz$ plane, rendering atomic sites within the same sublattice inequivalent, and is thus forbidden in $C_{3v}$ invariant systems. 

\begin{figure}
    \begin{centering}
    \includegraphics[width=0.75\textwidth]{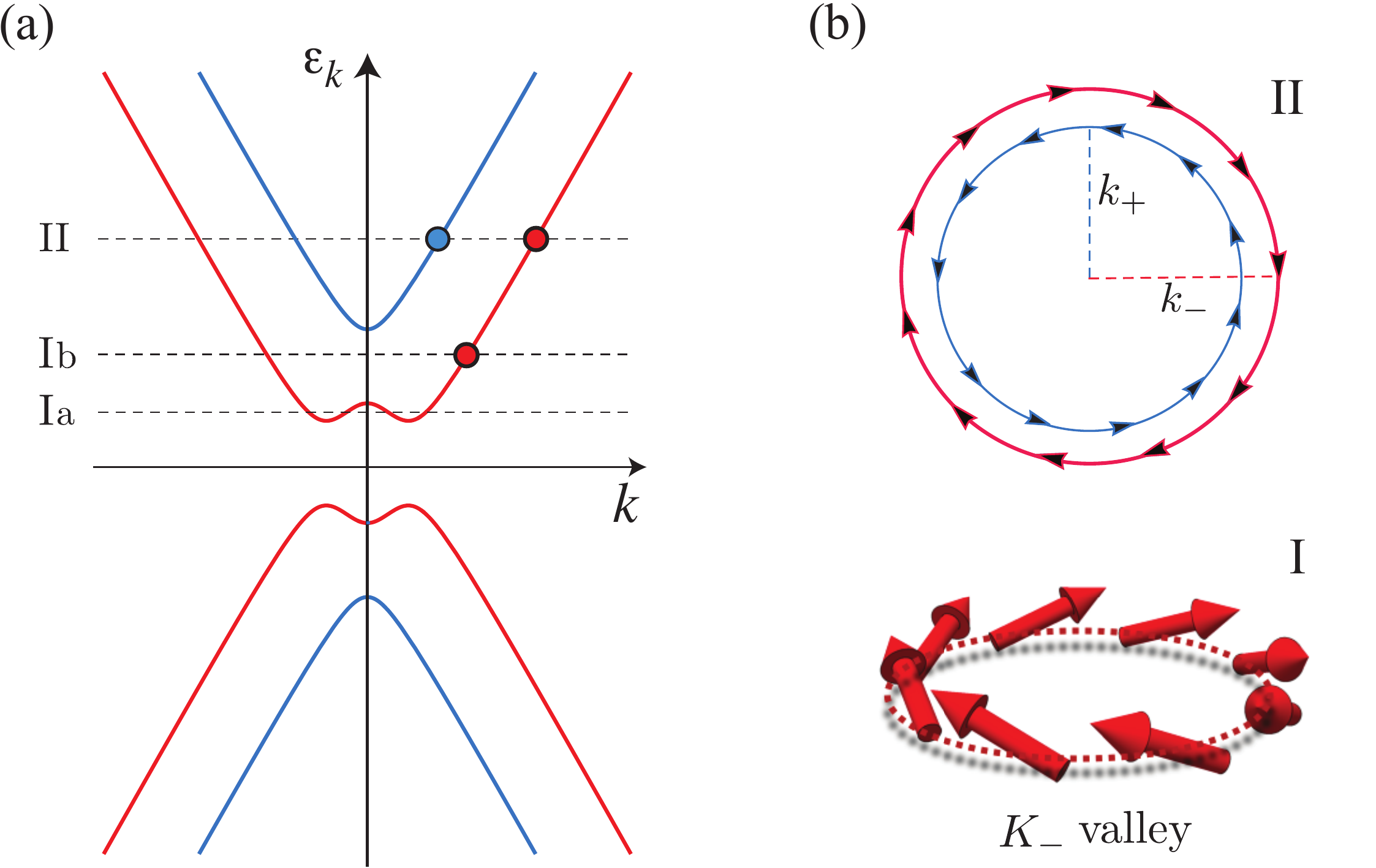}
    \par\end{centering}
    \caption{\label{fig:04}(a) Energy dispersion near the Fermi level.  Symmetry breaking SOC opens a spin-gap. A small Mexican hat develops in the regime I due to the interplay of spin-valley coupling and Rashba SOC. (b) Tangential winding of the spin texture in regimes I (3D view) and II (top view).}
\end{figure}

The full $C_{3v}$-invariant low-energy Hamiltonian in momentum space is therefore
\begin{equation}
    H_{\boldsymbol{k}}^{C_{3v}} = \hbar v \tau_{0} \, \boldsymbol{\sigma} \cdot \boldsymbol{k} \, s_{0} + m \tau_{z} \, \sigma_{z} \, s_{0} + \Delta_{\textrm{KM}} \tau_{0} \, \sigma_{z} \, s_{z} + \lambda_{\textrm{BR}} \tau_{0} \, \left( \sigma_{x} \, s_{y} - \sigma_{y} \, s_{x} \right) + \lambda_{\textrm{sv}} \tau_{z} \, \sigma_{0} \, s_{z}\,,
    \label{eq:C_3v_Ham_low_energy}
\end{equation}
where the above couplings have a simple correspondence to the spin-hoppings of the TB model {[}Eq.~(\ref{eq:C3v_invariant}){]} 
\begin{equation}
    \Delta_{\textrm{KM}} = \frac{\lambda_{A}+\lambda_{B}}{2}\,, \qquad \lambda_{\textrm{sv}} = \frac{\lambda_{A}-\lambda_{B}}{2} \,, \qquad \lambda_{\textrm{BR}} = \lambda\,.
    \label{eq:relation}
\end{equation}

The competition between the different spin-orbit energy scales in  Eq.~(\ref{eq:C_3v_Ham_low_energy}) influences the energy dispersion, while also dictating the topological properties and the spin structure of the eigenstates. Kane-Mele SOC ($\Delta_{\textrm{KM}}\tau_{0} \, \sigma_{z} \, s_{z}$) leads to spin-degenerate bands due to its spin-conserving nature, with the ability to drive the system into a topologically non-trivial phase when it dominates \parencite{Qian2014,Wu2018,Shi_19}. However, in the honeycomb monolayer systems of interest here, including TMDs and graphene-based vdW heterostructures, the Kane-Mele SOC is negligible. In contrast, the spin-valley coupling  ($\lambda_{\textrm{sv}}\tau_{z} \, \sigma_{0} \, s_{z}$), also known as valley-Zeeman interaction,  emerging from sublattice-resolved SOC is generally significant and yields spin-split bands within each valley. Typically, $\lambda_{\text{sv}}$ is of order 100 meV for semiconducting TMDs (intrinsic SOC) \parencite{Xiao_2012} and 1 meV for graphene-TMD heterostructures (proxomity-induced SOC) \parencite{Tiwari2022}. The Bychkov-Rashba term, $\lambda_{\textrm{BR}}\tau_{0}\left(\sigma_{x} \, s_{y}-\sigma_{y} \, s_{x}\right)$, arising at heterostructure interfaces, leads to spin admixture in the spin states, characterized by in-plane spin-momentum locking of the Bloch eigenstates. Similar to spin-valley SOC, this term results in spin-split bands distinguished by \textit{spin helicity} rather than simple spin. The magnitude of this coupling is typically on the order of 1 to 10 meV, depending on the TMD used  \parencite{Wang2019,Tiwari2022}. The presence of significant Rashba coupling explains the recently observed large charge-to-spin conversion via the ISGE at room temperature \parencite{Offidani_17}, as discussed later on in this article. Finally, it is worth noting that the sublattice-resolved spin-flip terms ($\lambda_{nn}^{A(B)}$) [see Eq.~(\ref{eq:C3v_invariant})] are absent in the low-energy Hamiltonian (i.e. they appear at the next order in the small-$\boldsymbol{k}$ expansion), and as such have little impact on the transport physics of 2D honeycomb layers.

The dispersion relation associated with Eq.~(\ref{eq:C_3v_Ham_low_energy}) consists of two pairs of spin-split Dirac bands. A typical energy dispersion relation for a graphene-TMD heterostructure with competing Rashba and spin-valley couplings is shown in Fig.\,\ref{fig:04}(a). The general electronic dispersion of $C_{3v}$ invariant systems can be readily seen as
\begin{equation}
    \varepsilon_{\zeta}(k) = \pm \sqrt{\hbar^{2} v_{}^{2} k^{2} + \Delta_{\zeta}^{2}(k)}\,,
    \label{eq:dispersion_relation}
\end{equation}
where $\Delta_{\text{KM}}$ has been neglected due to its inherently small nature compared to other SOC energy scales, $k\equiv|\boldsymbol{k}|$, $\zeta=\pm1$ is the \textit{spin-helicity} index, and $\Delta_{\zeta}^{2}(k)$ is the SOC-dependent mass term,
\begin{equation}
\Delta_{\zeta}^{2}(k)=m^{2}+\lambda_{\textrm{sv}}^{2}+2\lambda_{\textrm{BR}}^{2}+2\zeta\sqrt{\left(\lambda_{\textrm{BR}}^{2}-m\lambda_{\textrm{sv}}\right)^{2}+\hbar^{2}v^{2}k^{2}\left(\lambda_{\textrm{BR}}^{2}+\lambda_{\textrm{sv}}^{2}\right)}\,.
\label{eq:SOCmassterm}
\end{equation}
The spin texture of the eigenstates (see Fig. \ref{fig:04}(b)) can be cast in the following compact form
\begin{equation}
    \langle \boldsymbol{s} \rangle_{\zeta\tau\boldsymbol{k}}  = -\zeta \, \varrho(k) \, (\hat{k} \times \hat{z}) +  \, m_{\zeta \tau}^{z}(k) \, \hat{z}\,,
    \label{eq:spin_texture}
\end{equation}
where $\tau = \pm1$ is the  $K_{\pm}$ valley quantum number. The first term describes the spin winding of the electronic states generated by the BR effect \parencite{Rashba2009}. The second term is due to breaking of sublattice symmetry ($\lambda_{\textrm{sv}}\neq0$ or $m\neq0$) and tilts the spins in the $\hat{z}$ direction. Because of the Dirac nature of the charge carriers, the spin texture has a strong dependence upon the Fermi energy. In fact, the spins point fully out of the plane at the Dirac point, but  acquire an in-plane component as $k$ is increased. For $\hbar v k \gg \Delta_{\zeta}$, one finds $\rho(k)\simeq\cos\theta$ and $m_{\zeta\tau}^{z}(k)\simeq \sin\theta$, with $\theta=- \tau\,\arctan(\lambda_{\textrm{sv}}/\lambda_{\textrm{BR}})$. The tilting angle, $\theta$, has opposite signs in different valleys by virtue of time-reversal symmetry. Furthermore, one observes two distinct electronic regimes. For energies within the spin-gap (regimes Ia and Ib in Fig.\,\ref{fig:04}), the Fermi surface has a \emph{well-defined spin helicity}, a feature reminiscent of spin--momentum locking in topologically protected surface states \parencite{Schwab2011}. Consequently, near-optimal charge-to-spin conversion is observed inside the spin-gap via a large ISGE. In contrast, for energies outside the spin-gap (regime II in Fig.\,\ref{fig:04}), the spin helicity is no longer a well-defined concept, but the larger Fermi radius of the spin-majority band ($\zeta = -1$) nevertheless allows for a detectable ISGE  (i.e. while the current-induced spin polarization arising from the two sub-bands have opposite signs, they do not cancel each other). These special features of the electronic and spin structure of proximitized 2D materials are ultimately responsible for the efficient current-driven spin polarization supported by graphene-TMD heterostructures \parencite{Offidani_17}.

\section*{Relativistic Spin-Orbit Coupled Transport Phenomena}

\subsection*{Electronic Signatures}

As alluded to above, the presence of SOC in a material can drastically alter the transport properties of materials due to spin-charge inter-conversion  effects. One of the earliest signatures of SOC-affected transport was measured in magnesium films (i.e. 2DEGs) \parencite{Bergman1982,Sharvin1981}, where the electrical conductivity was seen to decrease upon the application of a magnetic field (i.e. a negative magneto-conductivity). This behavior can be traced back to the quantum interference between the many paths an electron can take when travelling through a disordered system. In the case of a material where SOC is absent, quantum interference leads to the coherent backscattering of electrons, thus yielding more localized states, and hence reduces the electrical conductivity in a phenomenon known as \textit{weak localization} (WL). The application of a magnetic field here destroys the coherence of the backscattered electrons and hence inhibits WL. Therefore, a positive magneto-conductivity is a clear signature of WL.

In contrast, materials that are SOC-active may exhibit the complete opposite of WL, \textit{weak anti-localization} (WAL). In this case, the paths of backscattered electrons interfere destructively to yield more delocalized states. This reversal in behavior can be understood to result from the spin DOF becoming an active participant in the Hamiltonian and scattering events. Here, the application of a magnetic field changes the interference of backscattered electrons from destructive to constructive, therefore decreasing the conductivity of a material. The natural signature of SOC-active materials is therefore a negative magneto-conductivity.

The role of localization in graphene, however, is more involved due to the presence of other spin-like DOFs, namely the sublattice and valley DOFs. In the absence of both SOC and inter-valley scattering (i.e. disorder is smooth on the lattice scale), graphene exhibits a WAL phase; the complete opposite of what is seen in 2DEGs. The manifestation of WAL in this system is a result of graphene's $\pi$ Berry phase, which precludes electrons from backscattering and thus prevents WL. Another interpretation can be found in considering the pseudospin (sublattice) DOF as an active participant in the Hamiltonian and scattering events, much in a similar manner to how spin becomes an active participant in SOC-active 2DEGs. Upon increasing the concentration of point defects and other short-range scatterers, such that inter-valley scattering no longer remains negligible, the disordered graphene system's localization phase reverses to exhibit a WL phase instead. Likewise, keeping intervalley scattering weak (i.e. intra-valley scattering dominates) and instead introducing a strong SOC also pushes graphene from a WAL to a WL phase. Finally, when both disorder and SOC are strong graphene moves back to WAL behavior \parencite{Sousa2022}. Several experiments on graphene-TMD and bilayer graphene-TMD have revealed WAL phases in these systems \parencite{Wang2016,Volkl2017,Yang2017,Wakamura2018,Amann2022}, indicating the presence of strong symmetry-breaking SOC. However, such observations do not provide a direct spectroscopic probe for the size of the various SOCs present. Only recently has the SOC of these systems been accessed directly \parencite{Wang2019,Tiwari2022}.

\subsection*{Spin Hall Effect}

While SOC might affect the electrical conductivity of graphene-based vdW heterostructures through quantum interference, more profound and exotic behavior can be observed when considering other forms of transport. The SHE is one such example of this, whereby the generation of a pure spin current driven solely by electric fields can be achieved without the need for magnetic fields. In pristine graphene -- owing to its negligible intrinsic SOC \parencite{Sichau_19} -- the SHE is absent. However, this scenario changes drastically in the presence of disorder-induced SOC \parencite{Ferreira_14,Balakrishnan2014,Milletari_16}. Here, the scattering of charge carriers from spin-orbit hot spots aligns spin and orbital angular momentum in opposite directions, resulting in the formation of transverse spin currents; the magnitude of the effect is characterized by the spin Hall angle, $\gamma$.

\begin{figure}
    \centering
    \includegraphics[width=0.9\columnwidth]{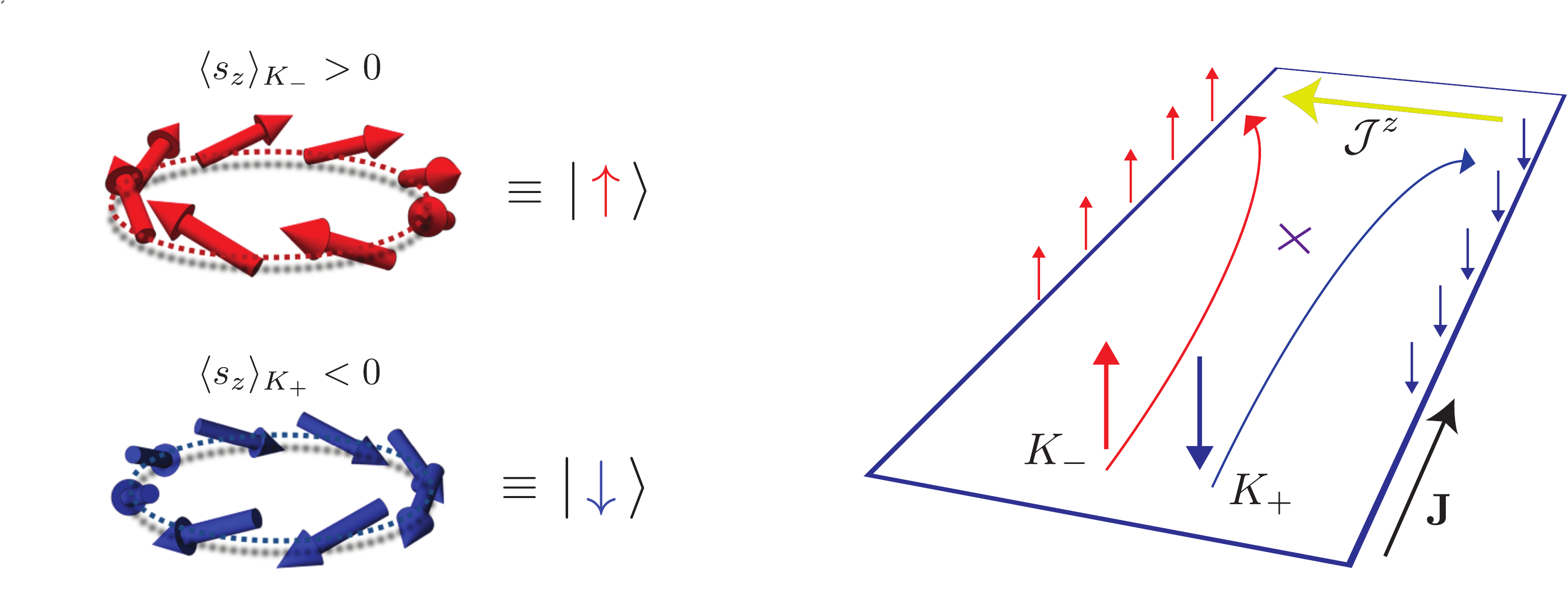}
    \caption{Left: Spin textures of spin-majority bands in each valley. The net out-of-plane spin polarization at each valley defines a quantization axis. Right: Spin accumulation at the edges of a thin film due to the SHE. Black arrow: electrical current, orange arrow: spin current, red/blue arrows: up/down spins. The electrons driven by the electrical current are scattered asymmetrically based upon their spin (skew scattering). Here, up spin electrons are scattered preferentially to the left, while down spins experience the opposite.}
    \label{fig:05}
\end{figure}

A direct consequence of the SHE in low-dimensional systems is the accumulation of spin at the system's boundaries. Specifically, thin films (2D materials) will accrue a collection of oppositely aligned spins at opposite edges, see Fig. \ref{fig:05}, due to the resulting spin current pumping up spins towards one boundary and down spins towards the opposite boundary. In thin wires (1D materials), the spin accumulation can be seen to wind around the wire's surface, much in a similar manner to how a magnetic field appears around a wire carring an electrical current. Reversing the direction of the applied electrical current reverses the spin accumulation in both cases mentioned here; the up and down spins now accumulate at the opposite boundary to which they had originally found themselves in thin films, and the direction of winding is reversed in thin wires.

Broadly speaking, the SHE arises in two forms: intrinsically and extrinsically. The former of these two is a direct result of the material's band structure, as opposed to being reliant upon external perturbations or extrinsic factors, such as disorder. This manifestation of the SHE can be understood in terms of an internal force experienced by the electrons travelling through the material due to the application of a charge current. This internal force is generated by the SOC experienced by the electrons, and hence is an effect embedded within the band structure of the system. An analogy can be drawn between this spin-orbit force driving the SHE and the Lorentz force responsible for the regular Hall effect. In contrast to the intrinsic SHE is the extrinsic contribution. Here, the generation of a spin current is driven by external mechanisms such as scattering from impurities. In this case, the SHE is the result of asymmetric scattering of electrons based upon their spin; up spins are scattered with a certain directional preference, while down spins are scattered with the opposite preference. This type of asymmetric scattering, known as \textit{skew scattering}, plays a central role in graphene-based vdW heterostructures \parencite{Milletari2017}.

To illustrate the importance of disorder-induced corrections to transport for these materials, consider the Hamiltonian for graphene with just Rashba coupling (i.e. $\lambda_{\text{sv}} = 0$ and neglecting $\Delta_{\text{KM}}$ given its small nature in most realistic scenarios). The carrier concentration is typically large enough to place the Fermi energy well above the Rashba pseudogap (i.e. in regime II of the band structure). Assuming only non-magnetic disorder is present, the system clearly lacks a well-defined quantization axis due to the lack of spin-valley coupling. Therefore, the SHE should not be observable simply because some form of SOC is present. With the inclusion of a spin-valley coupling it would then be natural to expect the possible occurrence an SHE as a quantization axis will now be well-defined.  However, if a theoretical analysis of the SHE for the minimal model without including disorder self-consistently is performed, it would lead one to believe that a finite SHE would exist in the absence of spin-valley coupling. This clearly emphasises the importance of including disorder in a fully self-consistent manner \parencite{Milletari2017,Milletari_16}.

Within the Kubo-Streda formalism of linear response \parencite{Streda1982,Crepieux2001}, the spin Hall conductivity may be written as $\sigma_{yx}^{\text{sH}} = \sigma_{yx}^{\text{sH,I}} + \sigma_{yx}^{\text{sH,II}}$, with \parencite{Milletari_16,Milletari2017}
\begin{subequations}
\begin{equation}
    \sigma_{yx}^{\text{sH,I}} = -\frac{1}{2\pi} \int_{-\infty}^{+\infty} d\omega \frac{df}{d\omega} \left\{ \text{Tr} \left[ \mathcal{J}_{y}^{z} G^{+}_{\omega} \widetilde{j}_{x} G^{-}_{\omega} \right] - \frac{1}{2} \text{Tr} \left[ \mathcal{J}_{y}^{z} G^{+}_{\omega} \widetilde{j}_{x} G^{+}_{\omega} + \mathcal{J}_{y}^{z} G^{-}_{\omega} \widetilde{j}_{x} G^{-}_{\omega} \right] \right\},
    \label{Kubo-Streda_type1}
\end{equation}
\begin{equation}
    \sigma_{yx}^{\text{sH,II}} = \frac{1}{2\pi} \int_{-\infty}^{+\infty} d\omega f(\omega) \text{Re} \left\{ \text{Tr} \left[ \mathcal{J}_{y}^{z} G^{+}_{\omega} \widetilde{j}_{x} \frac{\partial G^{+}_{\omega}}{\partial \omega} - \mathcal{J}_{y}^{z} \frac{\partial G^{+}_{\omega}}{\partial \omega} \widetilde{j}_{x} G^{+}_{\omega} \right] \right\},
    \label{Kubo-Streda_type2}
\end{equation}
\label{SHE_linear_Kubo}%
\end{subequations}
where $\mathcal{J}_{y}^{z}$ is the $z$-polarized spin current operator in the $y$-direction, $\widetilde{j}_{x}$ being the $x$ component of the disorder-renormalized electric current operator, $G^{\pm}_{\omega}$ are the retarded/advanced disorder-averaged Green's functions for electrons with energy $\omega$, $f(\omega)$ is the Fermi-Dirac distribution, and the trace is over momentum and all internal DOFs (spin, sublattice, and valley). Given the applied electric field, $\boldsymbol{E}$, generating an electrical current (assumed to be along the $x$-axis), the resulting spin Hall current  can then be determined using $\mathcal{J}_{y}^{\text{sH}} = \sigma_{yx}^{\text{sH}} E_{x}$.

The first term, $\sigma_{yx}^{\text{sH,I}}$, is sometimes referred to as the Fermi surface contribution, while the second term, $\sigma_{yx}^{\text{sH,II}}$, is similarly called the Fermi sea contribution. In weakly disordered systems, the type II contribution is higher order in the impurity density and hence may be neglected. Similar reasoning applies to the $G^{+}G^{+}$ and $G^{-}G^{-}$ terms in the type I contribution. Hence, it turns out that the leading order behavior is dictated solely by the cross term of $\sigma_{yx}^{\text{sH,I}}$. Clearly, the SHE in the metallic regime of disordered materials is essentially a Fermi surface property.

\begin{figure}
    \centering
    \includegraphics[width=\columnwidth]{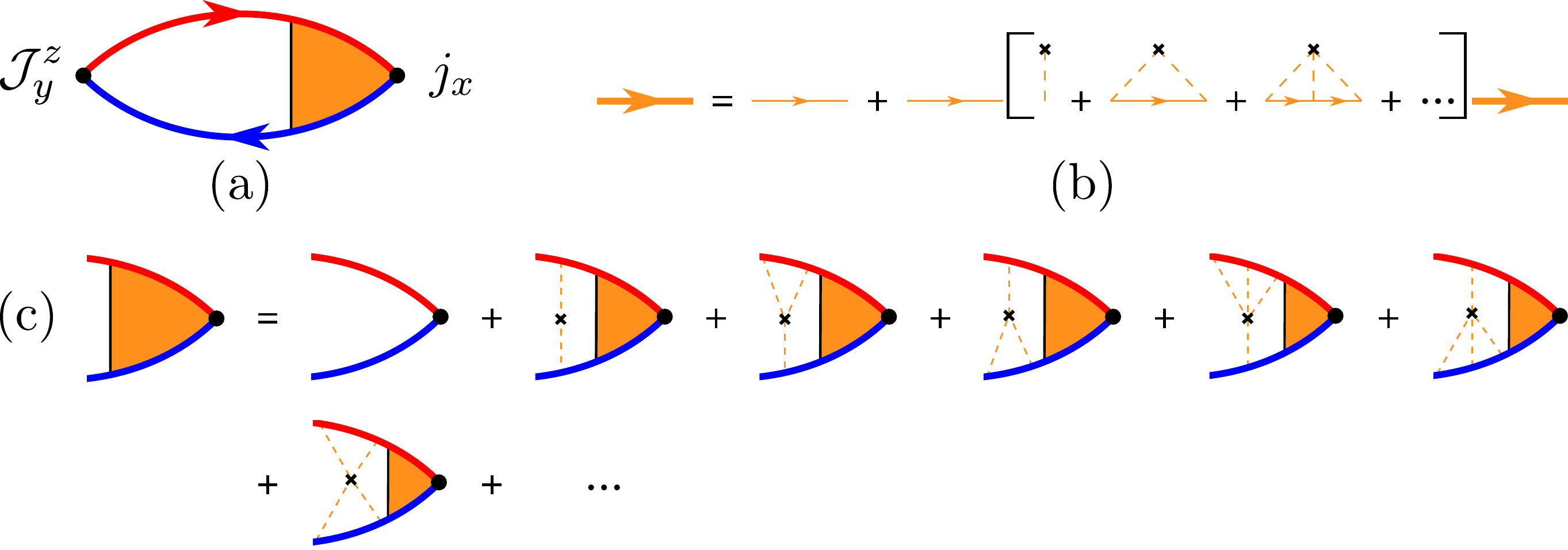}
    \caption{Diagrammatic representation of the cross term in $\sigma_{yx}^{\text{sH,I}}$ (a), the disorder-average Green's function (b), and vertex correction due to disorder (c). The thick red/blue lines denote retarded/advanced disorder-averaged Green's functions describing electron propagation, thick orange lines can be either retarded/advanced disorder-averaged electron Green's functions but must all be of the same type, thin orange lines indicate the free electron Green's functions, the black circles represent the current and spin current operators (vertices), black crosses represent an impurity, and the orange dashed lines denote scattering an electron scattering from the impurity.}
    \label{fig:06}
\end{figure}

A common representation that helps to visualize how disorder is included into linear response is that of Feynman diagrams. Figure \ref{fig:06} shows the dominant cross term in diagrammatic form, alongside the Dyson series describing the disorder-averaged Green's functions and the Bethe-Salpeter equation satisfied by the disorder-renormalized vertex. Note, the discussions and analysis here assume that only non-magnetic (spin-independent) scalar disorder is present. The dashed lines represent an electron (solid line) scattering from an impurity located at the cross. If $\sigma_{yx}^{\text{sH,I}}$ is evaluated without vertex corrections for the $\lambda_{\text{sv}} = 0$ case, one finds a non-vanishing result in complete contradiction to what is expected. However, including vertex corrections to any order in the number of scattering events, $\sigma_{yx}^{\text{sH,I}}$ can be seen to vanish, thus recovering the expected result for zero spin-valley coupling.

Interestingly, if one were to consider a Fermi energy located within the Rashba pseudogap (regime I of the band structure, $|\varepsilon| < 2 |\lambda_{\text{BR}}|$) they would also find a vanishing SHE in the absence of $\lambda_{\text{sv}}$. However, in this case, the type II contribution would no longer be sub-leading order and hence also needs accounting for (the cross term of the type I contribution remains the dominant part of $\sigma_{yx}^{\text{sH,I}}$). With this in mind, the computation of Eq. (\ref{SHE_linear_Kubo}) yields \parencite{Milletari2017}

\begin{equation}
    \sigma_{yx}^{\text{sH,I}} = \frac{e}{16\pi} \frac{|\varepsilon|}{\lambda_{\text{BR}}} = -\sigma_{yx}^{\text{sH,II}}.
\end{equation}
Hence, $\sigma_{yx}^{\text{sH}} = 0$ and so the SHE remains absent even in regime I due to the Fermi surface contribution being counteracted by off-surface processes.

Shifting focus to the experimentally relevant situation in which $\lambda_{\text{BR}}\neq 0$ and $\lambda_{\text{sv}} \neq 0$ while the SHE is expected to be observable (due to emergence of an effective spin quantization axis around each valley, see Fig. \ref{fig:05}), the role played by disorder changes drastically. If disorder is only accounted for within the Born approximation, where all processes involving three or more scatterings from a single impurity are neglected, then one finds Eq. (\ref{Kubo-Streda_type1}) yields a vanishing result. Only upon the inclusion of at least third order scattering events (at least three scatterings from a single impurity) into the vertex correction does one find a non-zero value for $\sigma_{yx}^{\text{sH,I}}$. By accounting for these higher order processes, a scattering event can now distinguish between left and right based upon the electron's spin along the well-defined quantization axis courtesy of the spin-valley coupling, i.e. skew scattering has been included. Therefore, not only are vertex corrections key to understanding the SHE completely, but so too are higher order scattering mechanisms allowing for the manifestation of the SHE. In other words, the extrinsic SHE due to scalar disorder effects in graphene-based vdW heterostructures is controlled entirely by skew scattering, which is always active provided the existence of a tilted BR-type spin texture at the Fermi level.

As a final note on intrinsic effects, couplings such as electron-phonon and electron-electron interactions, as well as structural defects such as ripples and shears in the individual layers of the heterostructure, are also considered as being intrinsic to the system. Consequently, the intrinsic SHE can potentially play an important role in clean systems where electron-phonon coupling dominates at room temperature, or in materials with sharp boundaries between structural domains. Just how major these effects are is still a current area of study, hence they shall not be covered in this article.

\begin{figure}
    \centering
    \includegraphics[width=0.5\columnwidth]{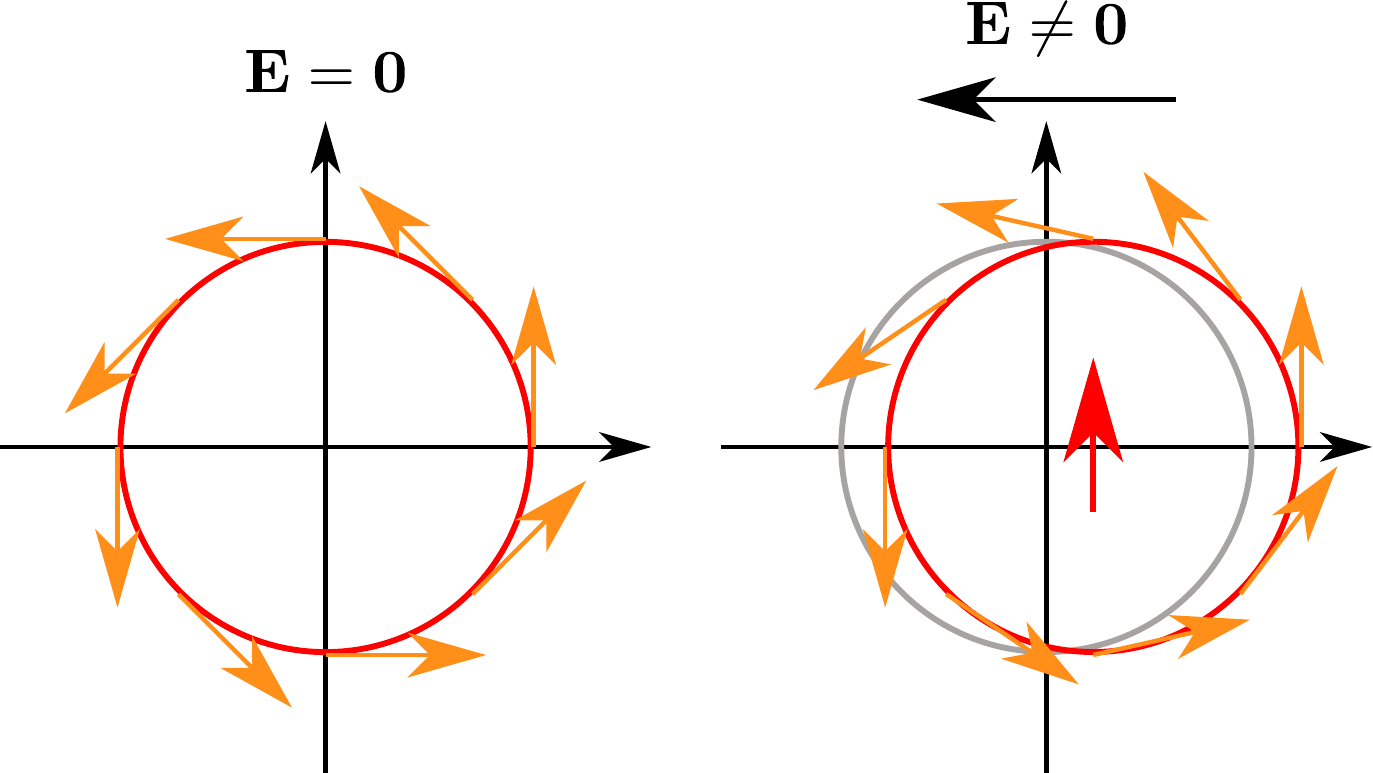}
    \caption{The manifestation of a non-equilibrium spin polarization with the application of an electric field is pictured above for a system with only Rashba SOC. Focus is placed on a single Fermi ring for ease of illustration. Left: Fermi ring for a system with zero current. The sum of electron spins (orange arrows) yields zero by symmetry. Right: an applied electric field shifts the Fermi ring from its original position (grey), with the electron spins on the Fermi surface rotating to maintain the spin-momentum locking typical of Rashba SOC, resulting in a non-zero spin polarization (red arrow). By inspection, all spins, except those on the $x$-axis, can be seen to rotate towards alignment with the $y$-axis. As a result, the net spin polarization is parallel to the $y$-axis in this case.}
    \label{fig:07} 
\end{figure}

\subsection*{Inverse Spin Galvanic Effect}

A natural partner to the SHE is the ISGE: the accumulation of spin upon application of an electric field without an associated spin current. Unlike the SHE, this spin accumulation manifests throughout the whole system. Rather, a non-trivial spin texture, facilitated by the interfacial breaking of inversion symmetry, is enough to allow for a non-zero spin polarisation when an electrical current is passed through the system. To demonstrate this, consider, once again, the minimal Dirac-Rashba Hamiltonian (only $\lambda_{\text{BR}} \neq 0$). It was shown above that the electron spins are locked in-plane and perpendicular to the electron momentum. In the absence of an electric field (i.e. in equilibrium), the Fermi rings forming the Fermi surface (two in regime II and Ia and one in regime Ib) are perfect circles around each Dirac point. Consequently, the electrons forming these rings yield a net spin polarization of zero (as expected of the nonmagnetic materials discussed here). However, when an electric field is introduced, these Fermi rings are shifted such that they are no longer rotationally symmetric about their respective Dirac points, see Fig. \ref{fig:07}. Therefore, the sum of the electron spins from each Fermi ring yield a non-zero spin polarization.

The direction of the resulting spin polarization is entirely dependent upon the system's spin texture. In the case of the Dirac-Rashba model above, the ISGE yields a spin polarization that is perpendicular to the applied electric field. The inclusion of a spin-valley coupling does not change the direction of the resulting spin polarization compared to the minimal Dirac-Rashba model. This is due to the out-of-plane component gained by the electron spins being opposite in sign between valleys (i.e. $\pm$ in the $K_{\pm}$ valleys). Hence, summing over the contribution from both valleys, one finds a vanishing $z$ component of spin accumulation. The role of the Kane-Mele-type SOC in these systems is negligible and hence does not contribute any meaningful changes to the spin texture either.

\begin{figure}
    \centering
    \includegraphics[width=0.5\columnwidth]{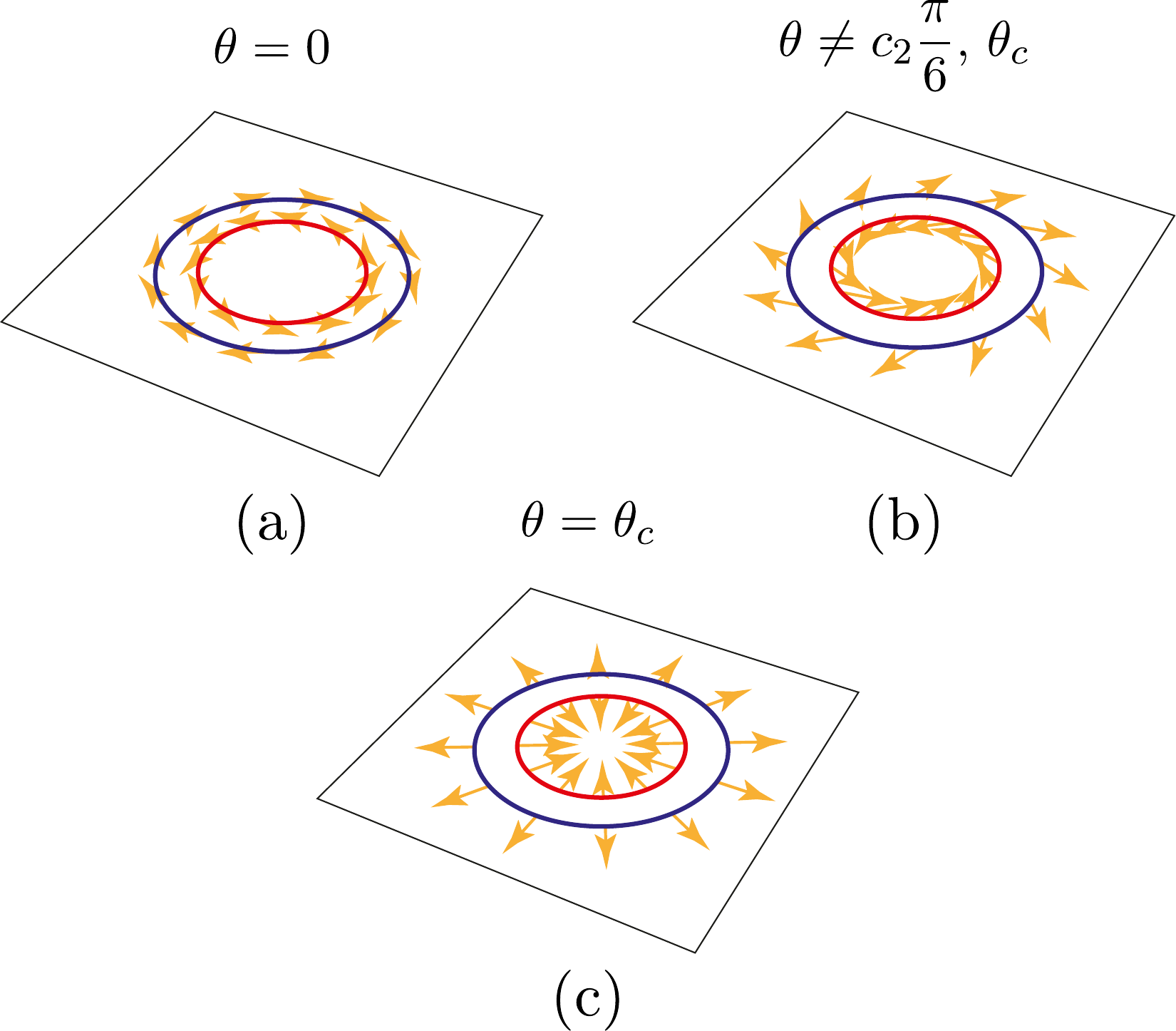}
    \caption{Introduction of a twist between the layers of a graphene-TMD vdW heterostructure can be seen to rotate the spin texture of the Fermi rings. (a): $\theta = 0$, all spin are locked in-plane and perpendicular to the momentum. (b): At arbitrary twist angles, $\theta \neq c_{2} \pi/6, \theta_{c}$, the electron spin are still fixed in-plane but not perpendicular of (anti-)parallel to the electron momentum. (c): At critical twist angles, $\theta = \theta_{c}$, the electron spins (anti-)align with their momentum.}
    \label{fig:08}
\end{figure}

An alternative method to manipulating the spin texture of these materials has been found through the introduction of a twist between a graphene monolayer and a TMD monolayer \parencite{Li2019,David2019,Peterfalvi2022,Veneri2022}. By introducing a rotational off-set between the two layers, both the Rashba and spin-valley couplings are affected and acquire a twist-angle, $\theta$, dependence. Not only are the magnitudes of these SOCs changed by twisting, but so too is the form of the Rashba term in the Hamiltonian. Upon the introduction of twisting, the spin-valley term maintains the same form as in Eq. (\ref{eq:C_3v_Ham_low_energy}) but with $\lambda_{\text{sv}}$ replaced by $\tilde{\lambda}_{\text{sv}}(\theta)$ ($\tilde{\lambda}_{\text{sv}}(0) = \lambda_{\text{sv}}$), and the Rashba term becomes
\begin{equation}
    H_{\text{R}}(\theta) = \tilde{\lambda}_{\text{BR}}(\theta) e^{i s_{z} \alpha_{\text{R}}(\theta)/2} (\sigma_{x} s_{y} - \sigma_{y}s_{x}) e^{-i s_{z} \alpha_{\text{R}}(\theta)/2},
 \label{eq:Rashba_phase}   
\end{equation}
where $\tilde{\lambda}_{\text{BR}}(0) = \lambda_{\text{BR}}$ and $\alpha_{\text{R}}(\theta)$ is known as the \textit{Rashba phase}. The exact way in which the Rashba coupling, spin-valley coupling, and Rashba phase vary with twist-angle depends heavily upon the material that graphene is paired with. A guaranteed property, however, is that $\alpha_{\text{R}}(\theta) = c_{1}\pi$ for $\theta = c_{2} \pi/6$ for $c_{1}, c_{2} \in \mathbb{Z}$. In any case, the spin texture will clearly be affected by the change in the Rashba coupling term's form. In fact, as the layers are twisted relative to one another, the electron spins can be seen to also rotate away from being locked perpendicular to their momenta. This rotation of the spin texture can be seen in Fig. \ref{fig:08}, where the out-of-plane component due to spin-valley coupling has been neglected for ease of illustration. It turns out that, for some materials, there exists a critical angle, $\theta_{c}$, where the spin texture is entirely radial (Fig. \ref{fig:08}c), sometimes referred to as a \textit{Weyl-type} or \textit{hedgehog} spin texture. Clearly, by being the perpendicular analog of the untwisted case, the resulting spin polarization will be perfectly collinear to the applied electric field. For any twist-angles away from $\theta = 0$, $\pm\theta_{c}$, $\pi/6$, the net spin polarization will be in-plane but neither perpendicular nor collinear to an applied electrical current. The value of $\theta_{c}$ is sensitive to the partner TMD used, atomic registry, strain distribution, and external perturbations (e.g. perpendicular electric field \parencite{Naimer_21}), and hence will vary significantly between materials. Consequently, meaningful predictions for $\theta_c$ are challenging to make. For example, graphene-WSe${}_{2}$ has been predicted to host a $\theta_{c} \approx 14^{\circ}$ (this prediction is based on an 11-band tight-binding model of the twisted heterostructures informed by both density functional theory calculations \parencite{Fang_15,Gmitra2016}  and available angle-resolved photoemission data \parencite{Pierucci_16,Nakamura_20}. In contrast, even the existence of a critical twist angle for graphene-MoS${}_{2}$ is difficult to ascertain given the variation in the reported material parameters of theory and experiment  \parencite{Peterfalvi2022}.

Defects and structural disorder are also expected to play a significant role. Of particular relevance is \textit{twist-angle disorder}; a type of spatial inhomogeneity that is ubiquitous in realistic systems \parencite{Uri_20}. Its impact on transport properties is expected to depend crucially on the comparative sizes of the spin diffusion length, $l_{s}$, and the twist-puddle size, $\xi$ (i.e. typical size of regions with a similar twist-angle). When $l_{s} \ll \xi$, the twist-angle disorder can be considered smooth and hence can be incorporated into  linear-response calculations through an appropriate averaging procedure (e.g. a Gaussian or box weighting could be applied \parencite{Veneri2022}). Though it is challenging to make first-principles predictions about the twist-angle behavior of the Rashba phase   for realistic systems, the  dependence of coupled spin-charge   transport phenomena on the Rashba phase can be reliably  studied   by replacing the standard Rashba SOC in the low-energy Hamiltonian of Eq. (\ref{eq:C_3v_Ham_low_energy}) with Eq. (\ref{eq:Rashba_phase}). In this case, the point group symmetry of the system, and therefore Hamiltonian, is once more reduced, going from $C_{3v}$ to $C_{3}$ for nontrivial twist angles.

To understand this from a quantitative perspective, one notes that the ISGE can be written mathematically as $S_{\alpha} = K_{\alpha\beta} E_{\beta}$ (assuming Einstein summation), where $K_{\alpha\beta}$ are the elements of the \textit{spin susceptibility} tensor. As in the SHE, the spin susceptibility tensor can be determined using linear response theory \parencite{Milletari_16,Milletari2017}, $K_{\alpha\beta} = K_{\alpha\beta}^{\text{I}} + K_{\alpha\beta}^{\text{II}}$,   
\begin{subequations}
\begin{equation}
    K_{\alpha\beta}^{\text{I}} = -\frac{1}{4\pi} \int_{-\infty}^{+\infty} d\omega \frac{df}{d\omega} \left\{ \text{Tr} \left[ s_{\alpha} G^{+}_{\omega} \widetilde{j}_{\beta} G^{-}_{\omega} \right] - \frac{1}{2} \text{Tr} \left[ s_{\alpha} G^{+}_{\omega} \widetilde{j}_{\beta} G^{+}_{\omega} + s_{\alpha} G^{-}_{\omega} \widetilde{j}_{\beta} G^{-}_{\omega} \right] \right\},
    \label{Kubo-Streda_type1_ISGE}
\end{equation}
\begin{equation}
\begin{split}
    K_{\alpha\beta}^{\text{II}} = \frac{1}{4\pi} \int_{-\infty}^{+\infty} d\omega f(\omega) \text{Re} \left\{ \text{Tr} \left[ s_{\alpha} G^{+}_{\omega} \widetilde{j}_{\beta} \frac{\partial G^{+}_{\omega}}{\partial \omega} - s_{\alpha} \frac{\partial G^{+}_{\omega}}{\partial \omega} \widetilde{j}_{\beta} G^{+}_{\omega} \right] \right\},
    \label{Kubo-Streda_type2_ISGE}
\end{split}
\end{equation}
\label{ISGE_linear_Kubo}%
\end{subequations}
where the Green's functions now contain the twisted form of the Rashba Hamiltonian. For disordered materials the $G^{+}G^{+}$ and $G^{-}G^{-}$ of the type I contribution, as well as the type II contribution, can once again be neglected. Consequently, the results for the in-plane components of the twisted spin susceptibility tensor can be related back to the untwisted ISGE, albeit with modified Rashba and spin-valley couplings,
\begin{equation}
\begin{split}
    K_{xx}(\tilde{\lambda}_{\text{BR}}(\theta),\tilde{\lambda}_{\text{sv}}(\theta);\theta) &= K_{yx}(\tilde{\lambda}_{\text{BR}}(\theta),\tilde{\lambda}_{\text{sv}}(\theta);0) \sin \alpha_{\text{R}}(\theta), \\
    K_{yx}(\tilde{\lambda}_{\text{BR}}(\theta),\tilde{\lambda}_{\text{sv}}(\theta);\theta) &= K_{yx}(\tilde{\lambda}_{\text{BR}}(\theta),\tilde{\lambda}_{\text{sv}}(\theta);0) \cos \alpha_{\text{R}}(\theta),
\label{eq:result_twisted}    
\end{split}
\end{equation}
\begin{equation}
    K_{yx}(\tilde{\lambda}_{\text{BR}},\tilde{\lambda}_{\text{sv}};0) = \frac{4ev\varepsilon}{\pi n u_{0}^{2}} \frac{\tilde{\lambda}_{\text{BR}}^{3} (\varepsilon^{2} + \tilde{\lambda}_{\text{sv}}^{2})}{\varepsilon^{4}(\tilde{\lambda}_{\text{BR}}^{2}+\tilde{\lambda}_{\text{sv}}^{2}) - \varepsilon^{2}\tilde{\lambda}_{\text{sv}}^{4} + 3\tilde{\lambda}_{\text{BR}}^{2}\tilde{\lambda}_{\text{sv}}^{4}},
\end{equation}
where $n$ is the impurity concentration and $u_{0}$ is the strength of the scalar impurities.  These results are strictly valid for perfectly aligned heterostructures. For systems with smooth twist-disorder landscapes ($l_{s} \ll \xi$), an intuitive approximate result can be obtained by taking the convolution of Eq. (\ref{eq:result_twisted}) with a suitable twist-disorder distribution function \parencite{Veneri2022}. For twist angles  $\theta\neq \theta_c, c_2\pi/6$, both in-plane components of the spin susceptibility tensor will be non-vanishing, thus yielding a non-trivial polarization ($S_{x},S_{y} \neq 0 $). At the critical twist angle the Rashba phase must vanish ($\alpha_{\text{R}}(\theta_{c}) = 0$), and hence a \textit{collinear Edelstein effect} (CEE) can be achieved, whereby the resulting spin polarization will be purely (anti-)parallel to the applied electrical current (c.f. Fig. \ref{fig:08}(c)). What makes these twisted graphene-TMD vdW heterostructures appealing is the ease with which the SOC and spin texture can be manipulated; only a simple twist is needed to adjust them. This sets these systems apart from 2DEGs, where both Rashba-type and Dresselhaus-type SOC \parencite{Trushin2007}. Control of these SOCs in 2DEGs is achieved via asymmetric doping and the tuning of quantum well widths \parencite{Ganichev2014}; a set of processes far more complicated and involved than simple twisting. For further details, the reader is referred to Ref. \parencite{Veneri2022} where the CEE phenomenon was predicted.

\begin{figure}
    \centering
    \includegraphics[width=0.7\columnwidth]{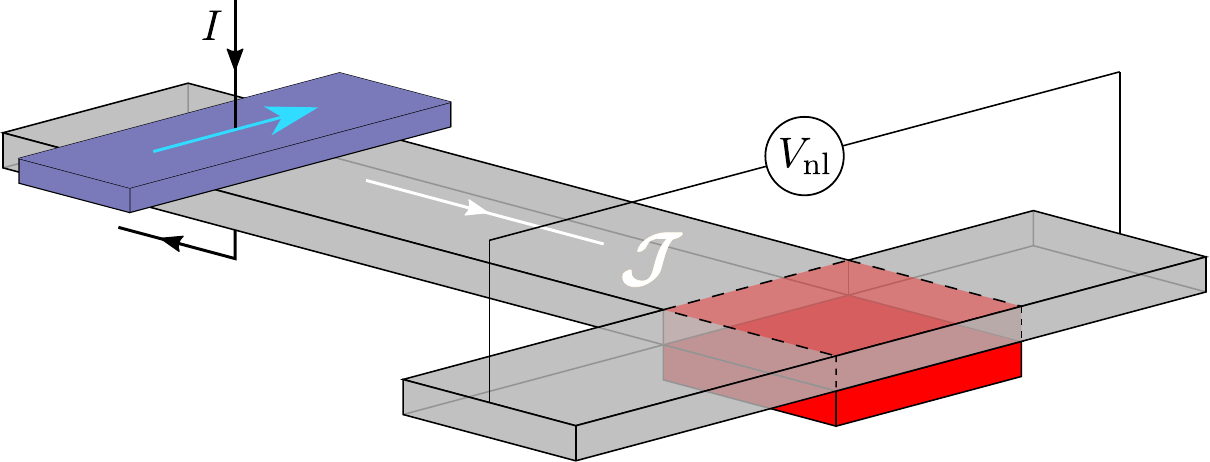}
    \caption{Graphene-based spin valve setup. An electrical current, $I$, passed through a ferromagnetic contact (blue box) injects a spin current into a graphene sheet. This spin current then travels along a graphene channel and into a region of enhanced SOC (dashed red region), due to proximity with a TMD (red box). This generates a charge accumulation in the arms of the T-junction due to the ISHE and SGE, which yields a non-local voltage in the steady state, $V_{\text{nl}}$, across the bar under the open circuit conditions.}
    \label{Spin_valve_setup}
\end{figure}

\subsection*{Observing Spin-Charge Interconversion}

To make use of graphene's large $l_{s}$ while also studying the effects of proximity-induced SOC, spin-valve setups are typically used to measure the ISHE and SGE (the inverse SHE and spin galvanic effect respectively). In this case, rather than trying to measure a spin current using an FM contact, an electrical current is measured instead by converting an injected spin current into an electrical current \parencite{Valenzuela_09}. A schematic of a spin-valve experiment is shown in Fig. \ref{Spin_valve_setup}.

Spin-valves operate by using a single FM contact to inject a spin current into graphene. This spin current is then able to flow along the isolated graphene channel until it reaches a T-junction, where it enters a region of high SOC. This region is no longer characterised by just monolayer graphene; instead, it now contains graphene layered on top of a TMD. As a result, the spin current entering this region undergoes the ISHE (due to any out-of-plane spin components) and generates an electrical current perpendicular to the injected spin current. Likewise, the electrons entering the high SOC region naturally also have a component of in-plane spin polarization and so are subject to the SGE, which also generates a perpendicular electrical current. Specifically, the component of the electron's spin parallel to the spin current generates an electrical current via the SGE, while the component of spin in-plane and perpendicular to the spin current yields an electrical current via the collinear SGE (if the material permits this process). The ensuing nonlocal resistance measured across the T-junction is therefore characterized by the spin-charge inter-conversion effects at the heart of modern spintronics.

In order to distinguish between the different electrical signals arising from the ISHE, SGE, and collinear SGE, a combination of magnetic fields must be used in conjunction with various FM orientations. The effect of applying a magnetic field to this setup is to cause spin precession about the applied field. The strength of the measured resistance in the T-junction will then depend on the magnetic field strength (how quickly the spins precess), and the length of the graphene channel (how long the spins have to precess). By combining the spin-valve measurements for various magnetic fields and FM orientations, the electrical signals associated to each of the aforementioned proximity-induced ISHE and SGE can be isolated by means of a simple symmetry analysis  \parencite{Cavill2020}. A set of recent spin-valve experiments have revealed graphene-TMD heterostructures to yield room-temperature nonlocal resistances of order 1-10 m$\Omega$ due to the ISHE and SGE \parencite{Safeer2019,Ghiasi2019,Benitez2020}. Alternatively, the Onsager-reciprocal phenomena, namely the SHE and ISGE, can be discerned by measuring the spin accumulation in the direction of the spin current at opposite sides of the high SOC region, as was done in the experiment of Ref. \parencite{Camosi_2022}. The latter approach requires a complex multi-terminal architecture but has the advantage that it permits isolation of the SHE and ISGE in situations for which the TMD is conducting and thus directly influences the spin-charge conversion processes (due to its high intrinsic SOC). 

Additionally, spin-valve measurements allow us to distinguish between conventional and anisotropic spin-charge conversion processes. An anisotropic SGE has been recently observed in graphene proximity-coupled to semimetallic (low-symmetry) TMDs   \parencite{Safee2019_2}. Likewise, an anisotropic ISGE characterized by the presence of spin polarization components parallel and orthogonal to the driving current  has been reported experimentally in \parencite{Camosi_2022}. However, an experimental demonstration exploiting twist-angle control in a graphene-semiconducting TMD heterostructures as described by Eq. (\ref{eq:result_twisted}) has yet to be achieved.

While spin-valve setups are ideally suited for studies of spin dynamics and spin-charge interconversion processes, they do not allow one to discern the size of the SOC present in a graphene-on-TMD heterostructure. To do this, measurement of the Shubnikov-de Hass (SdH) oscillations must be made. The reconstruction of the low-energy electronic structure from SdH data allows for the determination of the average SOC, $\bar{\lambda} = \sqrt{\lambda_{\text{BR}}^{2} + \lambda_{\text{sv}}^{2}}$ \parencite{Tiwari2022}. Experiments have revealed that the typical SOC present in these materials is $\bar{\lambda} \simeq 2.51$ meV. This average value is in accord to the predictions of microscopic theories for vdW heterostructures \parencite{Milletari2017,Offidani_17} regarding the observation of significant spin-charge interconversion efficiencies at room temperature, and is indeed compatible with the spin-valve measurements \parencite{Safeer2019,Ghiasi2019,Benitez2020,Camosi_2022,Li2020,Hoque2021}, thus demonstrating that proximity-induced SOC is responsible for the observed nonlocal resistances.

\section*{Summary}

This article has presented the role of graphene in modern spintronic devices, with a specific emphasis on how its partnership with other 2D materials can allow for bespoke systems with desirable electronic and spin transport properties. By constructing a low-energy theory of graphene, it becomes clear that transport phenomena in graphene-based systems will be dominated by the behavior of electrons around the Dirac points. From a physical point of view, one of the most striking features of isolated graphene is the description of its electrons as massless chiral Dirac fermions. Furthermore, the decoherence of spins in graphene occurs over large distances, allowing for long range spin transport, a direct result of the extremely weak intrinsic SOC (Kane-Mele) appearing naturally in graphene. However, despite its ability to host long range spin diffusion, graphene does not have a natural mechanism allowing for distinct spin control. This can ideally be achieved by pairing it with other materials that enhance its SOC.

In particular, the use of TMDs proximity-coupled to graphene gives rise to significant SOCs, of the order of meV, which allows meaningful spin currents and non-equilibirium spin polarizations, via the SHE and ISGE respectively, to manifest in the graphene layer. Consequently, electronic control of spin transport can be easily achieved and hence amalgamates well with current technological architectures. One major focus of spintronics is the implementation of the SHE and ISGE to generate large spin accumulations and polarizations that can be used to change the magnetization of a ferromagnet. In this case, the magnetization experiences a torque due to the SOC-related phenomena and so is referred to as a \textit{spin-orbit torque} (SOT). The use of SOTs in technology could allow for the realization of SOT-based magnetic random access memory (SOT-MRAM), removing our reliance upon volatile memory and thus reducing energy consumption. This route away from traditional RAM appears to bear much promise and remains as a current area of interest in research with many facets and subtleties still requiring further study.

As a final note, while not covered in this article, interface-induced magnetic exchange interaction is also becoming increasingly important in the field (e.g. as means to allow the manifestation of quantum anomalous Hall phases and spin-dependent Seebeck effect in graphene  \parencite{MEC_graphene_1,MEC_graphene_2}), and the reader is referred to the literature for further details. The same goes for high-temperature topologically nontrivial  phases of matter -- exhibiting technologically relevant phenomena like the quantum spin Hall effect \parencite{Wu2018} -- and the discovery of 2D vdW magnets \parencite{2Dmag_1,2Dmag_2,2Dmag_3}, which are likely to open up interesting avenues across many emergent fields, including topological quantum computation, spin-orbitronics, magnonics, and antiferromagnetic spintronics \parencite{2D_topological_1,2D_FM_spinorbitronics_1,2D_magnonics_1,2D_AF_spintronics_1}.

\section*{Acknowledgements}
The authors acknowledge support from the Royal Society through Grants No. URF/R/191021 (A.F.) and No. RF/ERE/210281 (A.F. and D.T.S.P.). We are indebted to Yue Wang and Robert A. Smith for helpful comments on the manuscript.

\newpage
\printbibliography
\end{document}